\documentclass[12pt]{article}
\usepackage{amsmath}
\usepackage{amsfonts}        
\usepackage{amssymb}
\usepackage{amsbsy}
\usepackage{mathrsfs}
\usepackage{epsfig}      
\usepackage{color}       

\usepackage[T1]{fontenc} 

\usepackage{graphicx}

\newlength{\TZ}
\DeclareFontFamily{OT1}{pzc}{}
\DeclareFontShape{OT1}{pzc}{m}{it}{<-> s * [1.200] pzcmi7t}{}
\DeclareMathAlphabet{\mathpzc}{OT1}{pzc}{m}{it}
\newcommand{\BEQ}{\begin{equation}}     
\newcommand{\BEA}{\begin{eqnarray}}
\newcommand{\BD}{\begin{displaymath}}
\newcommand{\EEQ}{\end{equation}}       
\newcommand{\EEA}{\end{eqnarray}}
\newcommand{\ED}{\end{displaymath}}

\newcommand{\eps}{\varepsilon}          
\newcommand{\vep}{\varepsilon}          
\newcommand{\vph}{\varphi}              
\newcommand{\D}{{\rm d}}                
\newcommand{\II}{{\rm i}}               
\newcommand{\demi}{\frac{1}{2}}         
\newcommand{\wit}[1]{\widetilde{#1}}    
\newcommand{\wht}[1]{\widehat{#1}}      

\renewcommand{\vec}[1]{\boldsymbol{#1}} 

\newcommand{\appsection}[2]{\setcounter{equation}{0}\setcounter{subsection}{0}
\section*{Appendix #1. #2}
\renewcommand{\theequation}{#1.\arabic{equation}}
              \renewcommand{\thesection}{#1}
              \renewcommand{\thefigure}{#1\arabic{figure}}\setcounter{figure}{0} }

\catcode`\@=11
\def\numberbysection{\@addtoreset{equation}{section}
        \def\theequation{\thesection.\arabic{equation}}}
\numberbysection                        

\definecolor{gruen}{rgb}{0,0.625,0}       
\definecolor{rot}{rgb}{0.75,0,0}          
\definecolor{blau}{rgb}{0,0,0.75}         
\definecolor{casta}{rgb}{0.45,0.20,0}     
\definecolor{gelb}{rgb}{0.825,0.725,0.0}  

\newcommand{\BLAU}[1]{\textcolor{black}{{#1}}}	    

\begin{document}

\begin{titlepage}

\vskip 1.5 cm
\begin{center}
{\Large \bf Physical ageing from generalised time-translation-invariance} 
\end{center}

\vskip 2.0 cm
\centerline{ {\bf Malte Henkel}$^{a,b}$
}
\vskip 0.5 cm
\begin{center}
$^a$Laboratoire de Physique et Chimie Th\'eoriques (CNRS UMR 7019),\\  Universit\'e de Lorraine Nancy,
B.P. 70239, F -- 54506 Vand{\oe}uvre l\`es Nancy Cedex, France\\~\\
$^b$Centro de F\'{i}sica Te\'{o}rica e Computacional, Universidade de Lisboa, \\Campo Grande, P--1749-016 Lisboa, Portugal\\~\\
\end{center}

\begin{abstract}
A generalised form of time-translation-invariance permits to re-derive the known generic phenomenology of ageing, 
which arises in \BLAU{classical} many-body systems after a quench from an initially disordered system to a temperature $T\leq T_c$, 
at or below the critical temperature $T_c$. 
Generalised time-translation-invariance is obtained, out of equilibrium, from a change of representation of the 
Lie algebra generators of the dynamical symmetries of scale-invariance and time-translation-invariance. 
Observable consequences include the algebraic form of the scaling functions for large arguments of the two-time auto-correlators and auto-responses, the equality of the
auto-correlation and the auto-response exponents $\lambda_C=\lambda_R$, the cross-over scaling form for an initially magnetised critical system 
and the explanation of a novel finite-size scaling if the auto-correlator or auto-response converge for large arguments $y=t/s\gg 1$ to a plateau. 
For global two-time correlators, the time-dependence involving the initial critical slip exponent $\Theta$ is confirmed and 
is generalised to all temperatures below criticality and to the global two-time response function, and their finite-size scaling is derived as well. 
This also includes the time-dependence of the squared global order-parameter. 
The celebrate Janssen-Schaub-Schmittmann scaling relation with the auto-correlation exponent is thereby extended to all temperatures below the critical temperature. 
A simple criterion on the relevance of non-linear terms in the stochastic equation of motion is derived, taking the dimensionality of couplings into account. 
Its applicability in a wide class of models is confirmed, for temperatures $T\leq T_c$. 
Relevance to experiments is also discussed. 
~\\~\\
\centerline{\textcolor{blau}{20$^{\rm th}$ of May, 2025 
}}
\end{abstract}

\vfill

\end{titlepage}

\setcounter{footnote}{0}



\section{Physical ageing: background} \label{sec:1}

\subsection{Introduction}

The precise description and the comprehension of the reasons underlying the observed phenomenology of non-equi\-li\-bri\-um dynamics continues to pose many 
challenges, both computational and conceptual. A prominent example is the dynamical behaviour of glassy systems \cite{Arce22,Bait22,Vinc24}. 
Much insight has been obtained in a classic series of experiments on the mechanical relaxation in glasses \cite{Stru78}, notably on the {\em ageing behaviour} in 
these systems. The set-up is typical in that the system was prepared in a high-temperature, molten state and then the ageing process was started 
by rapidly quenching the glass from its molten state to a temperature below the {\em glass-transition temperature}, where the glass solidifies. 
The sample was let to evolve up to the {\em waiting time} $s$ (up to the order of several years) 
when a mechanical stress was applied and the sample's response was measured at the {\em observation time} $t>s$. 
Remarkably, it was possible to identify several reproducible properties \cite{Stru78}, 
which are independent of the sample history or the detailed microscopic structure of the glass \cite{Zhai20,Arce22,Vinc24}. 
Prominent among these are features of dynamical scaling and that the form of the associated scaling function is independent of the material \cite{Stru78}. 
This is a first hint towards underlying dynamical symmetries, independently of many microscopic `details'. 
The same kind of experimental protocol can be applied to other non-equilibrium systems, notably to \BLAU{classical} magnetic systems (even without disorder) 
\cite{Bray94a,Bouc00,Cugl03,Cala05,Maze06,Odor08a,Puri09,Henk09,Henk10,Alba11,Taeu14,Giam16,Godr22} which are prepared in a
disordered high-temperature state before being quenched to either below $T<T_c$, with the critical temperature $T_c>0$ of the magnetic system, or else 
right onto the critical point $T=T_c$. Fixing the temperature $T$ after the quench and observing the system's behaviour leads to the same
qualitative features as seen before in glassy systems, including several distinct experimental confirmations \cite{Maso93,Alme21}. 
\BLAU{Implicitly, we shall concentrate throughout on classical systems and shall leave it to future work if an extension, and in which sense, 
to quantum dynamics is possible.}
We begin with the statement of the defining properties of physical ageing \cite{Stru78,Henk10}. 

\noindent
{\bf Definition:} {\it A many-body system is said to undergo {\em physical ageing} when its relaxation dynamics, from some initial state, obeys
the properties
\begin{enumerate}
\item slow relaxation dynamics {\rm (in contrast to {\em `fast relaxations'} of simple exponentials $e^{-t/\tau_{\rm r}}$ 
with a single and finite relaxation time $0<\tau_{\rm r}<\infty$)}
\item dynamical scaling holds {\rm (ageing phenomena are by definition dynamically scale-invariant and provide a natural basis for generalisations of this symmetry)} 
\item time-translation-invariance is absent
\end{enumerate}
}

%
\begin{figure}[t]
\includegraphics[width=.48\hsize]{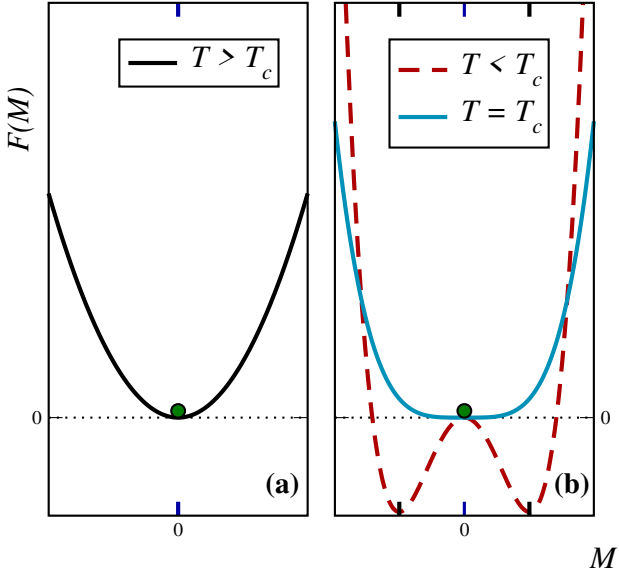} ~~~
\caption[fig1]{Schematic free energy of a simple ferromagnet (a) before a temperature quench and (b) after such a quench, either onto $T=T_c$ or else into $T<T_c$. 
The circle represents the system's disordered initial state. \label{fig1}}
\end{figure}
%
All three properties are required to specify the physical phenomenon we have in mind. \\
{\bf (1)} the requirement of {\em slow dynamics} distinguishes ageing systems from those in a disordered state with a single equilibrium state (when according 
to the general principles of thermodynamics, a fast relaxation within a finite characteristic time $\tau_{\rm r}$ towards that state should occur). 
Systems either at a critical point, or in a two-phase coexistence region, cannot undergo such a fast relaxation. 
To make this clear, we show in figure~\ref{fig1} the equilibrium free energy $F=F(M)$ (in the language of simple magnetic systems, where $M$ denotes the magnetisation). 
Figure~\ref{fig1}a shows the qualitative situation before the quench, with the system being prepared in a disordered initial state. After a rapid relaxation, 
the system's state is at the unique minimum of the free energy, with gaussian fluctuations around it. 
Now the system is quenched to either $T<T_c$ or $T=T_c$, see figure~\ref{fig1}b. 
Immediately after the quench, the system's state has not yet evolved but is no longer at a stable minimum. Below criticality ($T<T_c$), the system is
at an unstable maximum of $F(M)$, but because there are at least two equivalent minima of $F(M)$ available, the system cannot globally relax to just one of them. 
Microscopically, the system (rapidly) decomposes into many locally ordered clusters with a size $\ell(t)$ which grows with the time $t$ since the quench. 
Precisely at criticality ($T=T_c$), one has a more wide form of $F(M)$, because of critical non-gaussian fluctuations 
(see \cite{Joub08a,Joub08b} for an experimental equilibrium example at the Fr\'eedericksz transition in liquid crystals) and microscopically, correlated clusters
of linear size $\ell(t)$ will form.
In both cases $T\leq T_c$, a rapid relaxation towards equilibrium is impossible. 
Mathematically this is expressed through the relaxation time $\tau_{\rm r}$ becoming formally infinite. 
Hence only in the situations sketched in the right panel of figure~\ref{fig1} ageing is at all possible. \\
{\bf (2)} According to the requirement of {\em dynamical scaling}, there is a single time-dependent length scale $\ell=\ell(t)$. 
For large times $t\to\infty$, ageing is described by an increasing number of microscopic degrees of freedom and mean-field descriptions 
should become insufficient. 
To be specific, we shall restrict throughout to systems where $\ell(t)\sim t^{1/\mathpzc{z}}$ 
grows algebraically for large times,\footnote{Hence glassy systems where $\ell(t)\sim \bigl(\ln t\bigr)^{1/\psi}$ grows
logarithmically \cite{Fern19,Zhai20,Arce22} are not considered.} 
which defines the {\em dynamic exponent} $\mathpzc{z}$, whose value\footnote{An experimental example for non-gaussian fluctuations in phase-ordering is Cu$_3$Au, with 
$\frac{1}{\mathpzc{z}}=0.50(3)$ \cite{Shan92}.}   depends on whether $T<T_c$ or $T=T_c$.\\
{\bf (3)} The {\em absence of time-translation-invariance} means that ageing is an intrinsically 
{\em non-}equi\-li\-bri\-um phenomenon. This requirement serves to physically
distinguish ageing from critical dynamics {\em at} equilibrium, where time-translation-invariance holds. 

The precise meaning of `absence of time-translation-invariance' in the above definition will be the central topic of this work.  

The relaxation process is described via the time- and space-dependent {\em order-parameter} 
$\phi=\phi(t,\vec{r})$, conveniently coarse-grained such that an admitted continuum description becomes feasible. 
We shall normally admit that the equation of motion of $\phi$ is such that no conservation law is admitted (also referred to as {\em model-A dynamics})
and then speak of {\em phase-ordering kinetics} for a quench into $T<T_c$ and {\em non-equilibrium critical dynamics} for a quench onto $T=T_c$. 
Unless stated otherwise, we shall assume an initially disordered state with vanishing initial order-parameter 
$\bigl\langle\phi_{\rm init}(\vec{r})\bigr\rangle=\bigl\langle \phi(0,\vec{r})\bigr\rangle=0$, 
where the average $\bigl\langle \cdot\bigr\rangle$ is taken over the initial configurations and thermal histories. Then $\langle \phi(t,\vec{r})\rangle=0$ for all times $t>0$. 
The {\em two-time correlator} $C$ and the {\em two-time response} $R$ are defined\footnote{Spatial translation-invariance as well as spatial rotation-invariance such that 
$\vec{r}\mapsto r = |\vec{r}|$ will be admitted throughout, for notational simplicity.}  as \cite{Domi76,Jans76}
\BEQ \label{gl:1}
C(t,s;{r}) = \bigl\langle \phi(t,\vec{r})\phi(s,\vec{0})\bigr\rangle \;\; , \;\;
R(t,s;{r}) = \left. \frac{\delta \bigl\langle \phi(t,\vec{r})\bigr\rangle}{\delta h(s,\vec{0})}\right|_{h=0} 
=\bigl\langle \phi(t,\vec{r})\wit{\phi}(s,\vec{0})\bigr\rangle
\EEQ
where $\wit{\phi}(t,\vec{r})$ is the so-called {\em response scaling operator} from Janssen-de Dominicis non-equi\-li\-bri\-um field-theory \cite{Domi76,Jans76} 
see also (\ref{dynft}) below.\footnote{Response functions can also be computed directly from a specific correlator, without the need to specify explicitly a
non-vanishing external field \cite{Chat03,Ricc03,Chat11}.}  
Single-time correlators are included by letting $t=s$. In principle, one should be able to reconstruct the entire non-equilibrium physics from the
behaviour of these two-point (and all higher $n$-point) functions. 

\begin{figure}[tb]  
\includegraphics[width=.98\hsize]{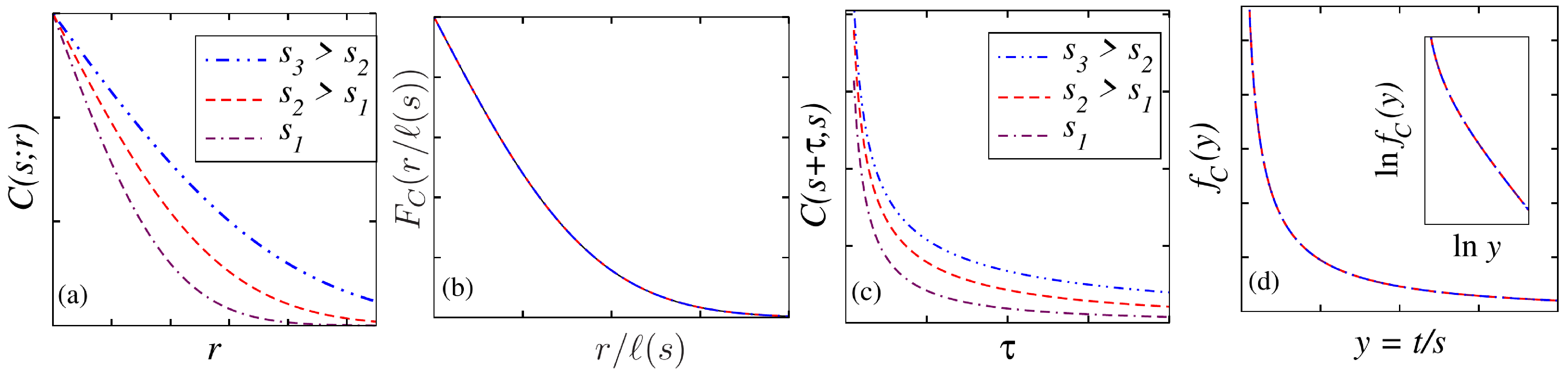}
\caption[fig2]{Physical ageing illustrated through the characteristic data collapses.
Panel (a) shows a typical behaviour of a single-time correlator $C(s;r)$ for different times
$s_1<s_2<s_3$, which collapse in (b) onto a single curve when distances $r=|\vec{r}|$
are measured in units of the dynamical length scale $\ell(s)$.
Similarly, panel (c) illustrates the two-time auto-correlator $C(s+\tau,s)$ in dependence of $\tau=t-s$,
for different waiting times $s_1<s_2<s_3$
which collapse in (d) when replotted as a function of $y=t/s$.
The inset shows the asymptotic power-law form $f_C(y)\sim y^{-\lambda/z}$.
\label{fig2} }
\end{figure}

The defining aspects of ageing are further illustrated in figure~\ref{fig2}, for definiteness for a quench into $T<T_c$. 
The {\em single-time correlator} $C(s;r) := C(s,s;r)$ and the {\em two-time auto-correlator}
$C(t,s) := C(t,s;0)$ are shown in figure~\ref{fig2}ac. First, we observe slow dynamics in the form that the system evolves more slowly when the time $s$ is increasing. 
Second, the correlators are seen to depend not only on the spatial distance $r$ or the time difference $\tau=t-s$, respectively, 
but also on the waiting time $s$ which makes it clear that the ageing process is not time-translation-invariant. Third, if the same
data are replotted as a function of $r/\ell(s)$ or $t/s$, respectively, a data collapse occurs, as is shown in figure~\ref{fig2}bd, and is evidence of dynamical scaling.

Usually, one now evokes dynamical scaling such that the two-point functions (\ref{gl:1}) are expected, for sufficiently large times $t,s\gg \tau_{\rm micro}$ and
$y=t/s>1$ (where $\tau_{\rm micro}$ is some reference time of microscopic size) to behave such that\footnote{Hence the time difference 
$\tau=t-s=s(y-1)\to \infty$ must be large. Finite values of $\tau$ are outside the scaling regime. 
The assumed scaling forms (\ref{gl:2}) explicitly exclude multi-scaling which might arise e.g. in {\em phase-separation kinetics} 
with a conserved order-parameter after a quench to $T<T_c$ \cite{Coni89,Cast99,Sire04}, as well as logarithmic sub-ageing \cite{Bert00,Dura17}.} 
\begin{subequations} \label{gl:intro} 
\BEQ \label{gl:2}
C(t,s;{r}) = s^{-b}   F_C\left( \frac{t}{s}; \frac{\bigl|\vec{r}\bigr|}{s^{1/\mathpzc{z}}}\right) \;\; , \;\;
R(t,s;{r}) = s^{-1-a} F_R\left( \frac{t}{s}; \frac{\bigl|\vec{r}\bigr|}{s^{1/\mathpzc{z}}}\right)
\EEQ
where $a,b$ are the {\em ageing exponents}. Their known values depend on whether $T<T_c$ or $T=T_c$. Finally, $F_{C,R}$ are scaling functions.
The dynamical scaling (\ref{gl:2}) is often referred to as {\em simple ageing}. In addition, it is often stipulated that the
{\em auto-correlator} $C(t,s) = C(t,s;0)$ and {\em auto-response} $R(t,s)=R(t,s;0)$, 
obtained when $\vec{r}=\vec{0}$, should have a simple scaling behaviour for $y\gg 1$ 
\BEQ \label{gl:3a}
f_C(y) = F_C(y,0) \sim y^{-\lambda_C/\mathpzc{z}} \;\; , \;\; f_R(y) = F_R(y,0) \sim y^{-\lambda_R/\mathpzc{z}}
\EEQ
where $\lambda_C,\lambda_R$ are the {\em auto-correlation} and {\em auto-response exponents}, respectively \cite{Huse89}. 
The power-law asymptotics of $f_C(y)$ for $y\gg 1$ is also indicated in figure~\ref{fig2}d. 
It is in general further admitted that one should have the exponent equality 
\BEQ \label{gl:3b}
\lambda_C=\lambda_R = \lambda 
\EEQ
\end{subequations} 
Both the scaling functions $F_C$, $F_R$ (or $f_C$, $f_R$) and {\it a fortiriori} the exponents $\mathpzc{z}$, $\lambda$ are believed to be universal, 
that is independent of the `microscopic details' of any given model. It is expected that the values of $\lambda$ and $\mathpzc{z}$ 
should be different for $T=T_c$ and $T<T_c$, respectively. All this is widely accepted {\it folklore} and has been confirmed countless times, either
in exactly solvable systems, numerical simulations or in experiments. 

For a totally disordered initial state and for model-A dynamics without any macroscopic conservation law, the following bounds are known \cite{Yeun96a,Fish88a,Henk15}
\BEQ \label{gl:lambda}
\lambda \geq \left\{ 
\begin{array}{ll} d/2                       & \mbox{\rm\small ~~;~ if $T<T_c$} \\
                  d-1+\eta/2 \geq d-2 +\eta & \mbox{\rm\small ~~;~ if $T=T_c$} 
\end{array} \right.
\EEQ
(with the known equilibrium exponent $\eta\leq 1$ \cite{Simo80}) which may serve as an useful consistency check on numerical or experimental results. 

Normally, it is not really discussed if these several assumptions (\ref{gl:intro}) are of an auxiliary nature, 
or if they are independent of each other, or even if it could be of interest to admit them only partially. 
It is one of the objectives of this work to give a generic discussion of these. 
Namely we shall show, that all the results (\ref{gl:intro}), and many other ones as well, 
follow from the unique hypothesis, besides the obvious dynamical scaling, of a generalised form of time-translation-invariance, combined
with standard scaling arguments. 
Indeed, we shall show in section~\ref{sec:3} that {\em the whole generic phenomenology of ageing \BLAU{in classical} systems can be derived from these two dynamical symmetries.}

\subsection{Preliminaries: the ageing exponents $b$ and $a$} 

As a preparation for the coming discussions, we first have to show how certain boundary or initial conditions, together
with dynamical scaling, can be used to fix the values of the ageing exponents $b$ and $a$ in (\ref{gl:2}). 

Consider a \BLAU{classical} many-body system, with some well-defined\footnote{The existence of conservation laws on $\phi$ should be
unimportant for what follows in this sub-section.} order parameter $\phi=\phi(t,\vec{r})$ after a quench
to a temperature $T\leq T_c$. First, the two-time correlation function should obey dynamical scaling \cite{Bray94a}
\BEQ \label{gl:C-skal}
C(t,s;{r}) = \kappa^{b\mathpzc{z}} C\bigl(\kappa^{\mathpzc{z}}t,\kappa^{\mathpzc{z}}s;\kappa {r}\bigr) 
               = s^{-b} F_C\left( \frac{t}{s};\frac{{r}}{s^{1/\mathpzc{z}}}\right)
\EEQ
as in (\ref{gl:2}). This must be supplemented by some external condition on the equal-time correlator. 
If the system is quenched into the phase-coexistence region where $T<T_c$, one expects, depending on the nature of the microscopic degrees of freedom 
\begin{subequations}
\begin{align}
C(s,s;{0}) \sim \left\{ \begin{array}{ll} 1            & \mbox{\rm ~discrete (e.g. Ising model)} \\
                                          M_{\rm eq}^2 & \mbox{\rm ~continuous} 
                         \end{array} \right.
\end{align}
whereas for a quench onto the critical point $T=T_c$, one should recover the large-distance critical equilibrium correlator 
\begin{align}
C_{\rm eq}({r}) = C(\infty,\infty;{r}) \sim \bigl|\vec{r}\bigr|^{-(d-2+\eta)} 
\end{align}
\end{subequations}
with the equilibrium critical exponent $\eta$. To reproduce this, the asymptotic scaling function $F_C$ in (\ref{gl:C-skal}), must satisfy for $u\ll 1$
\BEQ \label{gl:val-b}
F_C(1;u) \sim u^{-b\mathpzc{z}} \;\; ; \;\; 
b = \left\{ \begin{array}{ll}  0 & \mbox{\rm ~~if $T<T_c$} \\ {(d-2+\eta)}/{\mathpzc{z}} & \mbox{\rm ~~if $T=T_c$} \end{array} \right.
\EEQ
This furnishes the well-accepted value of the ageing exponent $b$ 
\cite{Bray94a,Cugl03,Henk10,Taeu14} from a boundary condition on dynamical scaling. 
In addition, for phase-ordering kinetics at $T<T_c$, and with a non-conserved order-parameter and short-ranged interactions, 
one has $\mathpzc{z}=2$ \cite{Bray94a,Bray94b}. More complicated results for $\mathpzc{z}$ hold for phase-separation and/or long-range interactions and need
not be recalled explicitly here. For non-equilibrium dynamics at $T=T_c$ the value of $\mathpzc{z}$ follows from detailed renormalisation-group
studies and is non-trivial \cite{Taeu14}. In disordered systems or spin glasses, $\mathpzc{z}=\mathpzc{z}(T)$ is temperature-dependent \cite{Bait18,Park10}.

Similarly, for a response function dynamical scaling gives 
\BEQ  \label{gl:R-skal}
R(t,s;{r}) = \kappa^{(1+a)\mathpzc{z}} R\bigl(\kappa^{\mathpzc{z}}t,\kappa^{\mathpzc{z}}s;\kappa {r}\bigr) 
= s^{-1-a} F_R\left(\frac{t}{s};\frac{{r}}{s^{1/\mathpzc{z}}}\right)
\EEQ
whose discussion requires a different kind of physical input. 
One must distinguish two classes of systems, namely 
\begin{enumerate}
\item {\bf class {\sc S}} with short-ranged interactions, in the sense that the equilibrium correlator 
      $C_{\rm eq}(|\vec{r}|)\sim e^{-|\vec{r}|/\mathpzc{r}_{\rm eq}}$ decays exponentially with $\mathpzc{r}_{\rm eq}$ finite. 
      
      Most systems quenched to $T<T_c$ are in this class (e.g. Ising or Potts models with $d\geq 2$). 
\item {\bf class {\sc L}} with long-ranged interactions, in the sense that $C_{\rm eq}(|\vec{r}|)\sim |\vec{r}|^{-(d-2+\eta)}$ decays algebraically. 

      All systems quenched onto $T=T_c>0$ are in this class. However, there are also systems quenched into $T<T_c$ which are in this class, for
      example the spherical model in $d>2$ dimensions with $\eta=0$. Systems such as the $1D$ Ising model with local interactions, quenched to $T=0$, 
      also belong to this class. 
\end{enumerate}
We now recall the arguments which lead to \cite{Henk03PP,Henk04PP,Henk10}
\BEQ \label{gl:val-a}
a = \left\{ \begin{array}{ll} 1/\mathpzc{z}              & \mbox{\rm ~~for class {\sc S}} \\ 
                              {(d-2+\eta)}/{\mathpzc{z}} & \mbox{\rm ~~for class {\sc L}} 
            \end{array} \right.
\EEQ
For disordered systems, there is evidence that (\ref{gl:val-a}) need not hold true \cite{Park10}. 

To see this, it is convenient to consider the response of the system to a harmonic perturbation
with angular frequency $\omega$ \cite{Bouc00}. The dissipative part of the alternating susceptibility is
\BEQ \label{gl:chi_pp}
\chi''(t;\omega) = \int_0^{t} \!\D u\: R(t,u) \sin\bigl(\omega(t-u)\bigr) 
= \chi''_{\rm eq}(\omega) + t^{-a} \chi''_2(\omega t) + \ldots
\EEQ
and this was decomposed into a stationary part $\chi''_{\rm eq}(\omega)$ and a scaling part $\chi''_2(\omega t)$. 

Now, for systems in class {\sc S}, one should decompose \cite{Bouc00} $\chi''(t;\omega)= \chi''_{\rm eq}(\omega) + \ell(t)^{-1} \chi''_{\rm age}(\omega t)$ 
into an equilibrium part and an ageing part.  The latter part is meant to represent the response of a single domain wall, 
where the domains have a linear size $\ell(t)\sim t^{1/\mathpzc{z}}$. Their volume is $\sim \ell(t)^d$ and their surface $\sim \ell(t)^{d-1}$, 
so that their density should scale as $1/\ell(t)$. Comparison with (\ref{gl:chi_pp}) gives the first result (\ref{gl:val-a}). 
For systems in class {\sc L}, which do not have well-defined domains with sharp walls, one rather should anticipate the decomposition
$\chi''(t;\omega) = \chi''_{\rm eq}(\omega) + \ell(t)^{-(d-2+\eta)} \chi''_{\rm age}(\omega t)$. This is because one expects a
quasi-static correlator $C(t,\vec{r})\approx C_{\rm qs}(\ell(t),\vec{r})\sim |\vec{r}|^{-(d-2+\eta)} \mathfrak{c}\bigl(|\vec{r}|/\ell(t)\bigr)$
with some scaling function $\mathfrak{c}$. The time-dependent contribution to the susceptibility per volume is expected to be
$\chi_{\rm qs} \sim V^{-1} \int_{V}\!\D \vec{r}\: C_{\rm qs}(\ell(t),\vec{r})\sim \ell(t)^{-(d-2+\eta)/\mathpzc{z}}$ which gives the second
part of (\ref{gl:val-a}). Many tests of this exist in the litt\'erature, including all known exactly solvable models, see  \cite{Henk10} and refs. therein.  

In the foregoing discussion, possible logarithmic factors have been systematically discarded. 

Eqs.~(\ref{gl:val-b},\ref{gl:val-a}) reproduce what is known since a long time for non-disordered \BLAU{classical} systems \cite{Godr02} and will be used as an input, where applicable, 
in the remainder of this paper.

This work is organised as follows. In section~\ref{sec:2}, we present our central postulate that non-equilibrium dynamics should admit a generalised form
of time-translation-invariance. In section~\ref{sec:3}, numerous phenomenological consequences of this postulate are derived and checked against the available evidence.
In section~\ref{sec:4} we investigate into a simple criterion which informs on the irrelevance of at least certain types of non-linearity for the long-time
dynamical symmetries in the equations of motion. We conclude in section~\ref{sec:5}. Several appendices contain technical details and calculations.

\section{Generalised time-translation-invariance} \label{sec:2}

The heart of this work resides in the following postulate on which representation of time-translations and dilatations to use. 

\noindent
{\bf Postulate:} {\it The Lie algebra generator $X_n^{\rm equi}$ of a time-space symmetry of an equilibrium system becomes a symmetry 
out-of-equilibrium by the change of representation}
\BEQ \label{gl:hyp}
X_n^{\rm equi} \mapsto X_n = e^{\xi \ln t} X_n^{\rm equi} e^{-\xi \ln t}
\EEQ
{\it where $\xi$ is a dimensionless parameter whose value characterises the scaling operator $\phi$ on which $X_n$ acts.} 

When applied to a dilatation generator $X_0^{\rm equi}=-t\partial_t - \frac{1}{\mathpzc{z}}r\partial_r -\delta$, the prescription (\ref{gl:hyp}) leads to a modified
effective scaling dimension $\delta_{\rm eff} = \delta-\xi$, since 
(spatial translation- and rotation-invariance are implicitly admitted, here and below, and are unchanged under (\ref{gl:hyp}))\footnote{\BLAU{This looks reminiscent to
an {\em equilibrium} critical-point scaling with an anomalous scaling dimension $\omega^*\geq 0$ of the vaccuum \cite{Fish73,Bake90}. 
A simple example occurs in the Yang-Lee singularity in an imaginary magnetic field \cite{Itzy86,Gehl91,Gao24}.}}  
\begin{subequations} \label{gl:Xgen}
\BEQ \label{gl:X0gen} 
X_0^{\rm equi} \mapsto X_0 = -t\partial_t - \frac{1}{\mathpzc{z}}r\partial_r - \bigl(\delta - \xi\bigr)
\EEQ
However, the time-translation generator $X_{-1}^{\rm equi}=-\partial_t$ becomes
\BEQ \label{gl:X-1gen}
X_{-1}^{\rm equi} \mapsto X_{-1} = -\partial_t + \frac{\xi}{t}
\EEQ
\end{subequations}
Whenever $\xi\ne 0$, time-translation-invariance is \BLAU{explicitly} broken, 
but this is achieved here through the choice of a different Lie algebra representation rather than changing the Lie algebra of dynamical symmetries by suppressing $X_{-1}$. 

The mathematical basis of a general change of representation (\ref{gl:hyp}) of a conformal algebra is summarised in appendix~A; 
here we focus on the change of dilatations and time-translations as specified in (\ref{gl:Xgen}). 

Heuristically, this might be argued as follows. {\em At} the critical point of an equilibrium phase transition, 
one of the spatial directions may be relabelled `time', while the other
ones are called `space'. Then one has time-translation and scale-invariance, obviously with $\xi=0$ and dynamical exponent $\mathpzc{z}=1$. 
This also remains valid when one discusses equilibrium critical dynamics, where
$t$ is now interpreted as a physical time but where time-translation-invariance and dynamical scaling are expected to hold, 
e.g. \cite{Card85}, and $\mathpzc{z}$ has a non-trivial value. 
If one now discusses {\em non-}equilibrium critical dynamics, still after a quench onto $T=T_c$ 
and which should obey dynamical scaling with a non-trivial $\mathpzc{z}$, e.g. \cite{Godr02,Cala05,Taeu14}, 
the lack of time-translation-invariance may be captured by letting $\xi\neq 0$ and the value of $\xi$ also
provides a measure of the distance with respect to equilibrium. Then, according to our postulate, 
the time-translation and dilatation generators should take the form (\ref{gl:Xgen}). 
Finally, one may also consider the case of coarsening dynamics after a quench into $T<T_c$, which maintains dynamical scaling \cite{Bray94a}  but is not
time-translation-invariant. We attempt to capture this by going over to the representation (\ref{gl:Xgen}) with $\xi\ne 0$. 

As a physical example for the relevance of $\xi$ for non-equilibrium dynamics, 
we consider briefly the kinetics of the exactly solvable {\em spherical model} \cite{Ronc78,Godr00b,Pico02,Henk23b}.
For $d=3$ it is in the same universality class as the $p=2$ spherical spin-glass \cite{Cugl95}. 
In this discussion, we anticipate results from  section~\ref{sec:4}. 
The model is formulated in terms of real-valued spin $S(t,\vec{r})\in\mathbb{R}$, subject to the constraint 
$\sum_{\vec{r}}\bigl\langle S^2(t,\vec{r})\bigr\rangle = {\cal N}$,
where $\cal N$ is the number of sites of the lattice. With the hamiltonian 
${\cal H}[S]=-\sum_{(\vec{r},\vec{r}')} S(t,\vec{r})S(t,\vec{r}') - \mu \sum_{\vec{r}} S^2(t,\vec{r})$ the
over-damped Langevin dynamics is given by 
\begin{subequations} \label{2.3}
\begin{align} \label{2.3a}
\partial_t S(t,\vec{r}) = -\frac{\delta {\cal H}[S]}{\delta S(t,\vec{r})} - \mathfrak{z}(t)S(t,\vec{r}) + \eta(t,\vec{r})
= \Delta_{\vec{r}} S(t,\vec{r}) - \frac{\digamma}{2\,t} S(t,\vec{r}) + \eta(t,\vec{r}) 
\end{align}
where $\eta$ is a thermal white noise and $\Delta_{\vec{r}}$ the spatial laplacian, in the continuum limit ${\cal N}\to\infty$. 
Herein, $\digamma=2\xi$ is obtained in section~\ref{sec:4.1} from the Lagrange multiplier $\mathfrak{z}(t)$ as follows. 
Defining $\ln g(t) := {2\int_0^t \!\D\tau\: \mathfrak{z}(\tau)}$, the spherical constraint becomes a Volterra integral equation for $g(t)$
\begin{align} \label{2.3b}
g(t) = A(t) + 2T \int_0^t \!\D \tau\: f(t-\tau) g(\tau)
\end{align}
\end{subequations}
with the temperature $T$ (for nearest-neighbour interactions, $f(t) = e^{-4dt}I_0(4t)^d$ with the modified Bessel function $I_0$). 
For a totally uncorrelated initial state $A(t)=f(t)$. In section~\ref{sec:4}, we shall obtain $1/t$-contributions, as in (\ref{2.3a}), 
in the equation of motion from our postulate (\ref{gl:hyp}).  
For non-equilibrium critical dynamics with $d>2$, after a quench onto $T=T_c(d)>0$ 
one finds for large times that $g(t)\sim t^{\digamma}$ with ${\digamma}=\min(d/2-2,0)$ \cite{Ronc78,Godr00b}. 
Hence, for $d>4$ the model is in the mean-field universality class and $\digamma=0$. But for $2<d<4$, fluctuation effects lead to $\digamma\ne 0$ and
require the change of representation (\ref{gl:Xgen}). Similarly, for coarsening after a quench to below $T<T_c(d)$, one has $\digamma=-d/2$ for all $d>2$. 
On the other hand, if one prepares the system at $T=T_c(d)$ {\em at critical equilibrium}, the model merely undergoes {\em equilibrium critical dynamics}. 
Equilibrium initial conditions are spatially long-ranged and modify $A(t)$ in (\ref{2.3b}) such that $g(t)=\mbox{\rm cste}.$ 
\cite{Pico02} and $\digamma=0$ throughout, for all $d>2$.  Since the equation of motion (\ref{2.3a}) is obtained from the change of representation (\ref{gl:hyp}) 
applied to the noisy diffusion equation of motion of the equilibrium spherical model, this illustrates the physical relevance of the
representations with $\xi\ne 0$, obtained via (\ref{gl:hyp}), 
for this example of non-equilibrium ageing phenomena.\footnote{This discussion only concerns the leading terms, for 
$t\to\infty$, of the solution $g(t)$ of (\ref{2.3b}), as the non-leading terms merely give corrections to scaling when inserted into (\ref{2.3a}).}

Will it be possible to extend the prescription (\ref{gl:hyp}) to larger symmetries, i.e. following ideas of local scale-invariance \cite{Henk94,Henk02,Henk10,Duva24}~?  
This must be considered cautiously. For example, in the most simple case $\mathpzc{z}=2$, 
a requirement of Galilei-covariance for correlators $C$, together with spatial translation-invariance, 
implies certain Bargman super-selection rules which lead to $C=0$ \cite{Pico04}. Non-trivial results may only be hoped for when co-variance under larger groups
is required for response functions $R$ only \cite{Henk02,Henk10}. From these, non-trivial correlators may be found by reducing them to certain response functions, 
taking into account detailed properties of the noise in the equations of motion \cite{Pico04}. We shall return to this question elsewhere. 
This work focuses on elaborating phenomenological consequences of the construction (\ref{gl:hyp}), restricted to time-translations and dynamical scaling, 
to permit a thorough test of this idea.\footnote{Generalised time-translations (\ref{gl:X-1gen}) 
arose implicitly in studies of local-scale-invariance where $\mathpzc{z}=2$ \cite{Henk06},
before being advocated as a special case of much more general representations \cite{Mini12} in the context of holographic geometry of ageing. See also appendix~A.} 

Our main use of dynamical symmetries will concern the scaling form of two-point functions. 
A scaling operator $\phi_a=\phi_a(t,\vec{r})$ is characterised by the
two parameters $(\delta_a,\xi_a)$. The co-variance of a {\em two-point function} (which may become a correlator or a response function, because of (\ref{gl:1}))  
\BEQ
\mathscr{C}=\mathscr{C}(t,s;\vec{r}) =\left\langle \phi_1(t,\vec{r})\phi_2(s,\vec{0})\right\rangle
\EEQ
under generalised time-translations (\ref{gl:X-1gen}) and dilatations (\ref{gl:X0gen}) leads to the conditions
\begin{subequations} \label{gl:inv}
\begin{align}
X_{-1} \mathscr{C} &= \left( -\partial_t - \partial_s + \frac{\xi_1}{t} + \frac{\xi_2}{s} \right) \mathscr{C} =  0 \label{gl:inv-1} \\
X_{0}  \mathscr{C} &= \left( -t\partial_t - s\partial_s - \frac{1}{\mathpzc{z}} r\partial_r -\delta_1 + \xi_1 - \delta_2 + \xi_2\right) \mathscr{C} = 0 
\label{gl:inv0}
\end{align}
\end{subequations}
Eq.~(\ref{gl:inv-1}) gives the meaning of `generalised time-translation-invariance' used in this work. 

In what follows, all physical applications will be consequences or adaptations of the following\\
{\bf Lemma.} {\it The general solution of the co-variance conditions (\ref{gl:inv}) is, for $t>s$ 
\BEQ \label{gl:loes}
\mathscr{C}(t,s;\vec{r}) = s^{-\delta_1-\delta_2+\xi_1+\xi_2} \left( \frac{t}{s} \right)^{\xi_1} \left( \frac{t}{s} -1 \right)^{-\delta_1-\delta_2} 
\mathscr{F}\left(\frac{r}{(t-s)^{1/\mathpzc{z}}}\right)
\EEQ
where $\mathscr{F}$ is an undetermined scaling function.} 

\noindent
{\bf Proof:} Clearly, (\ref{gl:loes}) solves both eqs.~(\ref{gl:inv-1},\ref{gl:inv0}). 
General theorems on linear partial differential equations \cite{Kamk79} guarantee the generality and the uniqueness of this form, since
a normalisation constant is absorbed into the undetermined function $\mathscr{F}$. \hfill ~q.e.d. 

\BLAU{The validity of the programme to be carried out in this paper will depend on the applicability of several auxiliary assumptions, which include 
\begin{enumerate}
\item[a)] simple ageing (\ref{gl:2}) with an algebraically growing length scale $\ell(t)\sim t^{1/\mathpzc{z}}$. 
\item[b)] all initial correlations will be assumed short-ranged. Otherwise the relation $\lambda_C=\lambda_R$, see (\ref{gl:3b}), may not hold 
\cite{Bray91,Bray94a,Pico02,Dutt08}. 
\item[c)] as postulated above, we shall restrict attention exclusively to the intertwining eq.~(\ref{gl:hyp}), which implies (\ref{gl:Xgen}).  
\end{enumerate}
Phenomenologically, we shall show in section~\ref{sec:3} that this choice is successful. 
It remains perfectly possible that in different physical situations, for instance at an upper critical dimension, or in multi-scaling, 
different forms must be used, instead of (\ref{gl:hyp}).
}

\section{Phenomenological consequences} \label{sec:3}

We now describe consequences of our postulate for the phenomenology of physical ageing of classical systems. Some of the results of this
section were announced before \cite{Henk25}. 

\subsection{Auto-correlation function}

The two-time auto-correlator $C(t,s) = \bigl\langle \phi(t,\vec{r})\phi(s,\vec{r})\bigr\rangle= \bigl\langle \phi(t,\vec{0})\phi(s,\vec{0})\bigr\rangle$ 
is built from two copies of the magnetic order-parameter $\phi$. Because of the 
identity of the scaling operators, we have
\BEQ \label{gl:3.1}
\delta = \delta_1 = \delta_2 \;\; , \;\; \xi = \xi_1 = \xi_2
\EEQ
The two-point function (\ref{gl:loes}) becomes, in the limit of large separations $y=t/s\gg 1$
\BEQ \label{gl:Cautoskal}
C(t,s) = s^{-2\delta +2\xi} \left( \frac{t}{s}\right)^{-2\delta+\xi} \mathscr{F}_C(0)
\EEQ
This does indeed agree with the expectations (\ref{gl:intro}) and in particular 
{proves the expected algebraic asymptotics (\ref{gl:3a}) of the scaling function $f_C(y)$}. We can identify
\begin{subequations} \label{18}
\begin{align}
\mbox{\rm if $T=T_c$:~~~} & b = 2(\delta-\xi) = \frac{d-2+\eta}{\mathpzc{z}} \;\; , \;\; 
\frac{\lambda_C}{\mathpzc{z}} = 2\delta - \xi = \xi + \frac{d-2+\eta}{\mathpzc{z}} \label{18a} \\
\mbox{\rm if $T<T_c$:~~} & b = 2(\delta-\xi) = 0 \;\;\hspace{1.5cm} , \;\; 
\frac{\lambda_C}{\mathpzc{z}} = 2\delta - \xi = \xi \label{18b}
\end{align}
\end{subequations}
where (\ref{gl:val-b}) was used. In summary, we have

\noindent
{\bf Proposition 1.} {\it The two-time auto-correlation obeys the scaling form $C(t,s) = s^{-b} f_C(t/s)$, 
where for large arguments one has for the scaling function $f_C$ the algebraic decay}
\BEQ
f_C(y) \stackrel{y\gg 1}{\simeq} f_{\infty,{\rm C}}\, y^{-\lambda_C/\mathpzc{z}} 
\EEQ
{\it where $\lambda_C/\mathpzc{z}$ is given by eqs.~(\ref{18}), for quenches onto $T=T_c$ and into $T<T_c$, 
respectively, and $f_{\infty,{\rm C}}$ is a constant.}

As discussed in appendix~B, it is {\em not} straightforward to include the single-time correlator $C(t;r)$ into this description. The problem of
its calculation is left open and we hope to return to it elsewhere. 

\subsection{Auto-response function}

The auto-response function $R(t,s) = \bigl\langle \phi(t,\vec{r})\wit{\phi}(s,\vec{r})\bigr\rangle$ 
can be written as a correlator of the order-parameter $\phi$ with the conjugate response
scaling operator $\wit{\phi}$ of Janssen-de Dominicis theory \cite{Domi76,Jans76}, which we shall discuss in more detail below, see eqs.~(\ref{dynft},\ref{dynact}). 
These scaling operators are characterised by the pairs $(\delta,\xi)$ and $(\wit{\delta},\wit{\xi})$ of parameters. 
Furthermore, we should eventually require responses to be co-variant under larger algebras of local scale-transformations, 
notably conformal transformations which make
up local scale-invariance \cite{Henk94,Henk02,Henk10}. If that is admissible, we have
\BEQ \label{3.5}
\delta = \delta_1 = \delta_2 =\wit{\delta} \;\; , \;\; \xi = \xi_1 \;\; , \;\; \wit{\xi} = \xi_2
\EEQ
The first identity (\ref{3.5}) is a consequence of local-scale-invariance. 
However, the two parameters $\xi$ and $\wit{\xi}$ remain independent.\footnote{{\em At} equilibrium, as discussed in section~\ref{sec:2}, 
both $\xi=\wit{\xi}=0$ vanish. 
Then $C=C(t-s)$ and $R=R(t-s)$ are time-translation-invariant from (\ref{gl:loes}) and obey the fluctuation-dissipation theorem.} 
Then, for $y=t/s\gg 1$ we have from (\ref{gl:loes})
\BEQ
R(t,s) = s^{-2\delta +\xi+\wit{\xi}} \left( \frac{t}{s}\right)^{-2\delta+\xi} \mathscr{F}_R(0)
\EEQ
once more in agreement with (\ref{gl:2},\ref{gl:3a}). We can identify, for both classes {\sc S} and {\sc L}
\BEQ \label{21}
1+a = 2\delta - \xi - \wit{\xi} \;\; ~~,~~ \;\; \frac{\lambda_R}{\mathpzc{z}} = 2\delta - \xi 
\EEQ
The results obtained so far can be summarised as follows.

\noindent
{\bf Proposition 2.} {\it For systems obeying generalised time-translation-invariance and dilatation-invariance, 
after a quench to $T\leq T_c$, 
the two-time auto-correlator $C(t,s)=s^{-b} f_C(t/s)$
and two-time auto-response $R(t,s)=s^{-1-a}f_R(t/s)$ have the properties:
\begin{enumerate}
\item For large arguments $y\gg 1$, both scaling functions are algebraic
\BEQ \label{3.8}
f_C(y) \simeq f_{\infty,{\rm C}}\, y^{-\lambda_C/\mathpzc{z}} \;\; ~~,~~ \;\; f_R(y) \simeq f_{\infty,{\rm R}}\, y^{-\lambda_R/\mathpzc{z}}
\EEQ
\item Comparison of (\ref{18}) and (\ref{21}) implies the exponent equality $\lambda=\lambda_C = \lambda_R$.
\end{enumerate}
}
Hence both well-established properties (\ref{gl:3a},\ref{gl:3b}) find their natural explanation as a consequence of generalised time-translation-invariance 
(\ref{gl:inv-1}), combined with dynamical scaling. They have been observed a countless number of times in the litt\'erature, see e.g. \cite{Henk10,Arce22,Vinc24} and 
references therein. The equality $\lambda_C=\lambda_R$ has been confirmed in field-theoretic studies to two-loop order both at $T=T_c$, see \cite{Cala05,Taeu14} 
for reviews, and also for $T<T_c$ \cite{Maze04}. 

In certain contexts, notably disordered systems, a good scaling collapse is only obtained if one works directly with the length scales $\ell(t)$ and $\ell(s)$. 
Phenomenologically, one would expect
\BD
C(t,s) = \ell(s)^{-B} \tilde{f}_C\left(\frac{\ell(t)}{\ell(s)}\right) \;\; , \;\; R(t,s) = \ell(s)^{-1-A} \tilde{f}_R\left(\frac{\ell(t)}{\ell(s)}\right)
\ED
where asymptotically $\tilde{f}_C(y) \sim y^{-\lambda_C}$ and $\tilde{f}_R(y) \sim y^{-\lambda_R}$ when $y\gg 1$. 
For this algebraic behaviour, there is well-documented evidence \cite{Park10,Park12}. 
Data for the disordered Ising model also support the equality $\lambda_C=\lambda_R$ \cite{Park10}, expected from Proposition~2. 
For simplicity, in this paper we shall not reformulate the generalised scaling postulate (\ref{gl:inv}) 
in terms of the length scales $\ell(t),\ell(s)$ but shall rather work throughout with the times $t,s$. 
Implicitly, we thereby restrict to the case $\ell(t)\sim t^{1/\mathpzc{z}}$ for large enough times, in the sense of \cite{Laka84}. 

In a different context, the understanding of the collapse dynamics of a $3D$ homopolymer, when the nature of the solvent is changed from good to bad by a temperature quench, 
has been greatly improved by borrowing concepts and methods from the domain coarsening of spin systems \cite{Maju16,Maju17,Chris17} and \cite{Maju20} for a review. 
In particular, the broad properties of physical ageing could be established. 
Power-law behaviour of both a length scale $\ell(t)\sim t^{1/\mathpzc{z}}$ and the two-time auto-correlator was found. 
While the dynamical exponent $\mathpzc{z}$ depends sensibly on the details
of the dynamics \cite{Chris17,Maju20}, the autocorrelation exponent $\lambda\simeq 1.25$ 
appears to be universal. Studies on the ageing in models of active matter \cite{Das17,Ditt22} 
or in Ising, spherical and voter models (the latter without detailed balance) with long-range interactions 
\cite{Cann01,Baum07,Chen12,Chris19,Chris20,Chris21,Corb21,Corb24a,Corb24b,Corb24c,Corb24d,Corb24e,Gess23,Gess24,Maju20,Muel24,Simo24} 
find the same qualitative characteristics. 

Besides the confirmation of the equality $\lambda_C=\lambda_R$ in numerous magnetic systems, ageing has also been studied in interface growth \cite{Bara95,Halp95,Taeu14}. 
It indeed holds true in the exactly solved Edwards-Wilkinson, Mullins-Herring \cite{Baum06b,Roethlein06a} and Arcetri
\cite{Henk15,Dura17}  models as well as in the $1D$ Kardar-Parisi-Zhang (KPZ) model \cite{Henk12}. 
However, in the $2D$ KPZ model, numerical simulation data have led to question the equality $\lambda_C\stackrel{?}{=}\lambda_R$ for a long time. 
Only recent data, with considerably improved numerical precision and effort, now seem to support $\lambda_C{=}\lambda_R$ \cite{Kell16,Kell17}.

Physical ageing has as well been investigated, and in particular the validity of the exponent equality $\lambda_C=\lambda_R$ confirmed, in several critical systems without
detailed balance, notably the contact process in several dimensions, non-equilibrium kinetic Ising and Potts models and also including experiments, 
see \cite{Baum07b,Enss04a,Hinrichsen06a,Odor06a,Odor08a,Ramasco04b,Takeuchi09a,Chat11,Boet18,Rich20}, 
spin-facilitated kinetically constrained models \cite{Maye03,Maye04,Maye04b,Maye06,Leon07}, 
several exactly solvable reaction-diffusion \cite{Baum05a,Durang09a,Durang11,Fort14} or voter models \cite{Corb24f} or  
the Kuramoto model of synchronisation \cite{Odor25}.    
 
Since it is often difficult to measure the noisy response function directly, it is very common to study integrated responses instead. Consider 
\begin{enumerate}
\item the {\em intermediate susceptibility}
\begin{subequations} \label{3.9} 
\begin{align}
\chi_{\rm int}(t,s) = \int_{s/2}^s \!\D u\: R(t,u) = \frac{1}{h} M_{\rm int}(t,s)
\end{align}
where $M_{\rm int}$ is the intermediate magnetisation and $h$ an external magnetic field, to be chosen small enough to remain in the linear-response regime. 
\item the {\em thermoremanent susceptibility}
\begin{align}
\chi_{\rm TRM}(t,s) = \int_{0}^s \!\D u\: R(t,u) = \frac{1}{h} M_{\rm TRM}(t,s)
\end{align}
where $M_{\rm TRM}$ is the thermoremanent magnetisation (often measured in spin glasses).
\end{subequations}
\end{enumerate}

\noindent
{\bf Corollary 1.} {\it For the two-time susceptibilities (\ref{3.9}), and if (\ref{gl:val-a}) holds, one has the asymptotics}
\begin{subequations} \label{3.10}
\begin{align}
\chi_{\rm int}(t,s) &= s^{-a} f_{\rm int}\left(\frac{t}{s}\right) \;\; \hspace{0.0cm}, \;\; 
f_{\rm int}(y) \stackrel{y\gg 1}{\simeq} f_{\infty,{\rm int}}\, y^{-\lambda/\mathpzc{z}} 
\label{3.10a} \\
\chi_{\rm TRM}(t,s) &= s^{-a} f_{\rm M}\left(\frac{t}{s}\right) \;\; \hspace{0.10cm}, \;\; 
f_{\rm M}(y) \stackrel{y\gg 1}{\simeq} f_{\infty,{\rm M}}\, y^{-\lambda/\mathpzc{z}} 
\label{3.10b}
\end{align}
\end{subequations}
{\it with constants $f_{\infty,{\rm int}}$ and $f_{\infty,{\rm M}}$.} 

We do not consider here any finite-time corrections to scaling. 
The details of the proof are given in appendix~C.\footnote{The also often-measured
zero-field-cooled susceptibility $\chi_{\rm ZFC}(t,s) = \int_s^t \!\D u\: R(t,u)$ may contain a dominant contribution from the upper limit of integration 
(notably in ferromagnets quenched to $T<T_c$), besides a contribution analogous to (\ref{3.10}). See \cite{Henk03PP,Henk04PP,Henk10} for a detailed discussion.} 
The power-law behaviour (\ref{3.10b}) of $f_{\rm M}(y)$ for large $y\gg 1$ 
is indeed a very long-standing observation, notably for spin glasses \cite{Gran87}, has been at the focus of many theoretical studies \cite{Bouc92,Bouc94} 
and is by now standard {\em folklore} \cite{Arce22,Vinc24}, including but not limited to spin-glass physics. 
It is very satisfying that it comes out here as a natural consequence of the generalised time-translation-invariance (\ref{gl:inv-1}).  

\begin{table}[tb]
\begin{center}
\begin{tabular}{|l|cl|l|} \hline 
material                                  & ~condition~          & ~$X_{\infty}$~   & ~Ref. \\ \hline
inner ear hair cell                       & ~morphology~         & ~$\simeq 0.25$~  & ~\cite{2Martin01a,2Dinis12a} \\[0.08cm]
CdCr$_{1.7}$In$_{0.3}$S$_4$               & $T< T_c$             & ~$0.2 - 0.4$~    & ~\cite{Herisson02a,Herisson04a}~ \\[0.08cm]
PMMA                                      & $\phi\simeq \phi_c$  & ~$0.43(6)$~      & ~\cite{Wang06a} \\[0.08cm]
liquid crystal 5CB                        & $\eps\simeq 0$       & ~$\simeq 0.31$~  & ~\cite{Joubaud09a} \\[0.09cm] 
$\gamma$Fe$_2$O$_3$ {nano}-particles      & $T=0.3 T_g$          & ~$\simeq 0.15$~  & ~\cite{2Komatsu11a,2Nakamae14a} \\[0.00cm]
                                          & $T=0.4 T_g$          & ~$\simeq 0.26$~  & \\[0.09cm]
collo\"{\i}d in active bath~              &                      & ~$0.40(4)$~      & ~\cite{2Maggi17a} \\[0.08cm]\hline
\end{tabular}
\caption[tab1]{Some experimentally measured values of the
limit fluctuation-dissipation ratio $X_{\infty}$, defined in (\ref{3.11}), in spin glasses, a liquid crystal, {nano}-particles and biological matter.
\label{tab:1} }
\end{center}
\end{table}

Another interesting consequence concerns the {\em fluctuation-dissipation ratio} $X(t,s)$ and the limit fluctuation-dissipation ratio $X_{\infty}$, 
defined as \cite{Godr00b,Godr02}
\BEQ \label{3.11}
X_{\infty} := \lim_{s\to\infty} \left( \lim_{t\to\infty} X(t, s) \right) 
= \lim_{s\to\infty} \left( \lim_{t\to\infty} T R(t,s) \left( \frac{\partial C(t,s)}{\partial s} \right)^{-1} \right)
\EEQ
where the order of the limits is important. At equilibrium, the fluctuation-dissipation theorem implies $X_{\infty}=1$. 
Table~\ref{tab:1} gives a short list of experimentally measured finite values of $X_{\infty}$ in non-equilibrium systems. There
are considerable differences between the values of $X_{\infty}$ in the various systems considered. 

\noindent
{\bf Corollary 2.} {\it For quenches onto $T=T_c >0$, the limit fluctuation-dissipation ratio $X_{\infty}$ (\ref{3.11}) 
is a finite (usually positive, but not always) constant.}

\noindent {\bf Proof:} from the scaling (\ref{3.8}) one has for $t/s\gg 1$ that 
\BD
\frac{\partial C(t,s)}{\partial s} \simeq s^{-1-b} \left(\frac{\lambda_C}{\mathpzc{z}}-b\right) f_{\infty,{\rm C}} \left( \frac{t}{s}\right)^{-\lambda_C/\mathpzc{z}}
\ED
and by re-using (\ref{3.8}) we have, again for $t/s\gg 1$ and the definition (\ref{3.11}) 
\BD
X(t,s) \simeq \frac{T_c f_{\infty,{\rm R}}}{f_{\infty,{\rm C}}} \frac{\mathpzc{z}}{\lambda_C -\mathpzc{z}b}\, 
              s^{b-a} \left( \frac{t}{s} \right)^{(\lambda_C-\lambda_R)/\mathpzc{z}}
\ED
Proposition~2 states $\lambda_C=\lambda_R$ and comparing eqs.~(\ref{gl:val-b},\ref{gl:val-a}) for a quench onto $T=T_c$ gives $b=a$. 
Hence the limit fluctuation-dissipation ratio $X_{\infty}$ is a finite constant. With the bound (\ref{gl:lambda}) and $\eta\leq 1$ 
it is positive (we assume $f_{\infty,{\rm C}}>0$ and $f_{\infty,{\rm R}}>0$ to be positive). 
\hfill q.e.d. 

At criticality, $X_{\infty}$ is thought to be universal \cite{Godr00b,Godr02}, 
and this has been confirmed many times in the litt\'erature \cite{Maye03,Chat04,Cala05,Garr05,Henk04a,Maye06,Leon07,Henk10,Chat11,Bait17,Vodr22,Corb24f,Odor25}. 
It is satisfying to see the critical $X_{\infty}$ to come out to be a finite 
constant and to have it related to universal exponents and to the amplitude ratio $T_c\, f_{\infty,{\rm R}}/f_{\infty,{\rm C}}$. 
The finiteness of the experimentally measured values of $X_{\infty}$, see table~\ref{tab:1}, is an indirect confirmation of $\lambda_C=\lambda_R$ and $b=a$. 
In kinetically constrained systems, $X_{\infty}$ is sometimes found to be negative \cite{Maye04,Maye06,Leon07}. 

In the past, one had often argued that the finiteness of $X_{\infty}$ at criticality implies the identities $b=a$ and $\lambda_C=\lambda_R$. 
However, that kind of reasoning is impossible for quenches into $T<T_c$, since then $X_{\infty}=0$. 
It had not been possible to understand why one should have $\lambda_C=\lambda_R$ in that latter case, but we now see from Proposition~2 that it must hold true.

\subsection{Finite-size effects in correlators}

Consider an ageing system which is placed into a finite volume, for definiteness of a hyper-cubic geometry
$\overbrace{N\times \cdots \times N}^{\mbox{\rm $d^*$ factors}}\times\overbrace{\infty\times\cdots\times\infty}^{\mbox{\rm $d-d^*$ factors}}$ 
such that in the finite directions there is a linear size $N$. Such finite volumes may for example arise in experiments from the samples having a natural graininess or
else by deliberate restriction of the dimensions of the sample. If $d=d^*$, one speaks of a {\em fully finite system}. 
In fully finite systems, finite-size effects will lead to the ageing process not going on forever, 
but rather to an interrupted ageing, e.g. \cite{Bouc94,Joh00,Buis03,Joub08a,Theu15,Zhai17,Kenn18,Barbier21,Arce22,Vinc24},  
after some time $t_{\rm stop}$ which will depend on $N$.\footnote{Even in systems known to undergo simple ageing, finite-size effects may create the illusion of an
effective but spurious sub-ageing behaviour \cite{Chris25}.}  
With respect to the spatially infinite systems considered up to now, 
how will the shape of the auto-correlator $C(t,s;N^{-1})$ be modified through finite-size effects~? 

\begin{figure}[t]
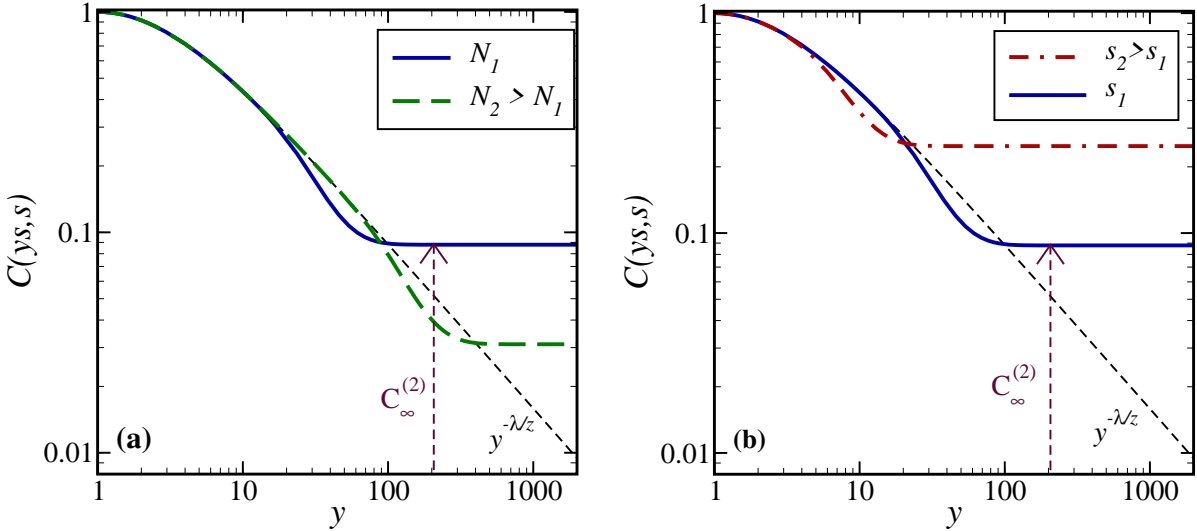

\includegraphics[width=.45\hsize]{TD-symmetrie-CN.eps} ~~~\includegraphics[width=.45\hsize]{TD-symmetrie-Cs.eps} ~~~~~~
\caption[fig3]{Finite-size scaling of the two-time auto-correlator $C(ys,s;N^{-1})$ in a fully finite volume (for phase-ordering with $b=0$). 
(a) Change of the plateau height $C_{\infty}^{(2)}$ with varying $N$ for fixed waiting time $s$. 
(b) Change of the plateau height $C_{\infty}^{(2)}$ with varying $s$ for fixed size $N$.
The dashed line gives the infinite-system auto-correlator with its asymptotic power-law $C(ys,s)\sim y^{-\lambda/\mathpzc{z}}$.\label{fig3}}
\end{figure}

Figure~\ref{fig3} illustrates, for a fully finite system and for phase-ordering kinetics at $T<T_c$ (hence $b=0$) the behaviour of the two-time 
auto-correlator $C(ys,s;N^{-1})$ with respect to the infinite-size auto-correlator $C(ys,s) \sim y^{-\lambda/\mathpzc{z}}$ 
(dashed line) and in a situation when $\ell(s)\ll N$ is kept fixed but $\ell(t)$ is allowed to grow up to saturation, see also (\ref{3.16}) below. 
While for $N\to\infty$ a power-law behaviour is observed, the saturation of the larger length $\ell(t)$ leads to deviations from this form
at times when $\ell(t)\approx N$. As long as $y=t/s$ is not yet too large, one still has a data collapse indicative of the ageing of an effectively infinite-sized systems 
but when $y$ is increased there is no longer
a data collapse and the finite-size effects lead to an {\em interruption} of ageing. 
One observes that for larger values of $y$ the auto-correlator decays more rapidly than in the infinite system before saturating at a plateau of height 
\BEQ
C_{\infty}^{(2)} = \lim_{y\to\infty} C\left(ys,s;\frac{1}{N}\right)
\EEQ
In figure~\ref{fig3}a it is shown that $C_{\infty}^{(2)}$ decreases with $N$ when $s$ kept fixed but
in contrast figure~\ref{fig3}b illustrates the increase of $C_{\infty}^{(2)}$ with $s$ when $N$ is kept fixed, 
at least as long as the waiting times are still small enough such that $\ell(s)\ll N$. 
Can one understand this behaviour in terms of a scaling description based on (\ref{gl:hyp})~? 
How does the plateau height $C_{\infty}^{(2)}=C_{\infty}^{(2)}(s;N^{-1})$, found in fully finite systems, depend on either $N$ or the waiting time $s$~? 
And what can eventually be said when $d^*<d$ and the system is not fully finite~? 
%
%

According to the theory of finite-size scaling \cite{Fish71,Barb83,Suzuki77} at equilibrium,\footnote{In general, the notion of {\em finite-size scaling} refers to the scaling 
behaviour of a system with respect to the linear extent $N$. In studies of phase separation, it has been useful to analyse the scaling of infinite-size systems 
with respect to the time-dependent length scale $\ell(t)$, which is certainly finite for a finite time $t$ \cite{Maju10,Maju11,Midya14}. 
This technique has been adapted in studies of  the collapse in homopolymers, see \cite{Maju17} for a pedagogical introduction. 
It has become common to emphasise the analogy with equilibrium and to call the technique `finite-size scaling'.}  
in a finite volume the inverse size $1/N$ becomes a further relevant variable. Therefore, the two-time auto-correlator (with spatial 
separation $\vec{r}=\vec{0}$) should obey the two co-variance conditions 
\begin{subequations} \label{gl:etf}
\begin{align}
X_{-1} {C} &= \left( -\partial_t  - \partial_s  + \frac{\xi}{t}                                   + \frac{\xi}{s} \right) {C} =  0 
\label{gl:etf-1} \\
X_{0}  {C} &= \left( -t\partial_t - s\partial_s + \frac{1}{\mathpzc{z}} \frac{1}{N}\partial_{1/N} -2\bigl(\delta - \xi\bigr)\right) {C} = 0 
\label{gl:etf0}
\end{align}
\end{subequations}
which have the unique solution (re-using eqs.~(\ref{18}) and (\ref{gl:3b}), of course)  for $t\gg s$
\BEQ \label{25}
C\left(t,s;\frac{1}{N}\right)  = s^{-b} \left( \frac{t}{s} \right)^{-\lambda/\mathpzc{z}} \mathscr{F}_C\left( \frac{N}{t^{1/\mathpzc{z}}} \right)
\EEQ
This form of non-equilibrium finite-size scaling shows that the behaviour of the auto-correlator depends on the ratio $N/\ell(t)$ whereas the waiting-time $s$ in (\ref{25}) 
is always kept small enough such that $\ell(s)\ll N$, the linear extent of the finite directions. 
Clearly, the form of the finite-size scaling function $\mathscr{F}_C(u)$ will depend on the boundary conditions in the finite directions. 

For a fully finite system $d^*=d$. The finite-size scaling function $\mathscr{F}_C(u)$, 
of the single argument $u=N t^{-1/\mathpzc{z}}$, controls the interpolation between the finite-size and the infinite-size systems. Clearly, for large arguments 
$\mathscr{F}_C(\infty)=\mbox{\rm cste.}$, such that the infinite-size behaviour (\ref{gl:Cautoskal}) is recovered. 
On the other hand, for $u\ll 1$, one should have $\mathscr{F}_C(u)\sim u^{-\lambda}$ such that the plateau is indeed 
$y$-independent, as suggested by figure~\ref{fig3}. This implies for the plateau height $C_{\infty}^{(2)} \sim s^{\lambda/\mathpzc{z}-b} N^{-\lambda}$. 
The requirement (\ref{gl:etf-1}) of generalised time-translation-invariance is essential to reduce the scaling function $\mathscr{F}_C$ to a function of
a single variable on which the traditional finite-size scaling arguments \cite{Fish71,Barb83,Suzuki77} 
can be applied. Heuristic arguments based on having $\ell(s)\ll N$ and the growing or saturation of $\ell(t)$ give the same result \cite{Henk25b,Wark25}. \\

\noindent
{\bf Proposition 3:} {\it For quenches to $T\leq T_c$ in a fully finite system of linear size $N$, and if the waiting time satisfies $s\ll N^{\mathpzc{z}}$, 
the two-time auto-correlator $C$ obeys the finite-size scaling (\ref{25}).
For large times it converges to a plateau, of height $C_{\infty}^{(2)}$, which obeys the scalings
\begin{subequations}\label{gl:etf-plateau}
\begin{align}
\mbox{\it if the waiting time $s$ is fixed:~~~}  &  ~C_{\infty}^{(2)} \sim N^{-\lambda}               \label{gl:etf-plateauN} \\
\mbox{\it if the system size  $N$  is fixed:~~~} &  ~C_{\infty}^{(2)} \sim s^{\lambda/\mathpzc{z} -b} \label{gl:etf-plateaus}
\end{align}
\end{subequations}
where the values (\ref{18}) must be used for the exponent $b$.}

It is understood that the values of $b$ and of $\lambda/\mathpzc{z}$ are those of the infinite system. 

In experiments,\footnote{Especially the recent ones studying the phase-ordering in liquid crystals \cite{Alme21}.} 
since the sizes of the grains come from the sample preparation and might be difficult to control, 
it should be more easy to keep $N$ fixed and look for the variation of the plateau with $s$,
using (\ref{gl:etf-plateaus}). Confirmations of (\ref{gl:etf-plateau}) in exactly solvable models exist for the spherical model quenched to 
$T<T_c$ and with $2<d<4$ \cite{Henk23} and the $1D$ Glauber-Ising model quenched to $T=0$ \cite{Henk25b}. 
For infinite sizes, the fully connected $p=2$ spherical spin glass is in the same dynamical universality class as the $3D$ spherical model \cite{Cugl95}, but
on a finite lattice and times $t\gtrsim t_{\rm cross}\sim N^{2/3}$ and $s\ll t_{\rm cross}$ the noise-averaged two-time auto-correlator, after a quench to $T=0$, 
converges to a plateau in agreement with (\ref{gl:etf-plateau}) and new values $\mathpzc{z}=\frac{2}{3}$ and $\lambda=\demi$ \cite{Fydo15}, see also appendix~D. 
Tests of (\ref{gl:etf-plateau}) in the $2D$ \cite{Wark25} and $3D$ \cite{Sast25} Glauber-Ising model for $T<T_c$ are forthcoming. 
Similarly, saturation effects were seen in the Kuramoto model long ago \cite{Ioni14}. 
As a computational tool, it is still unclear if (\ref{gl:etf-plateau}) will produce more precise results than more traditional methods. 

What does figure~\ref{fig3} further imply for practical calculations/measurements trying to determine $\lambda$ in a given system~? 
Up to now, calculations used the infinite-system asymptotics (\ref{3.8}), tried to study sufficiently large lattices to avoid any finite-size effects 
and used a double logarithmic plot of $f_C(y)$ for as large values of $y$ as possible. 
Theoretically, you do {\em not} expect a perfect power-law -- at least when $\mathpzc{z}=2$ as it
occurs in phase-ordering \cite{Bray94b}, as the leading finite-$y$ correction 
$f_C(y)\simeq f_{\infty,{\rm C}}\,y^{-\lambda/2}\bigl( 1 - A/y\bigr)$ with $A\geq d-\lambda$
can be derived from Local Scale-Invariance \cite{Chris20}. 
This might entice people into interpreting a slight curvature in the data as an effect of this leading correction and not,
as suggested by figure~\ref{fig3}, as the possible onset of the cross-over towards to plateau $C_{\infty}^{(2)}$. 
Not recognising this carries the risk of systematically over-estimating $\lambda/\mathpzc{z}$. 
Therefore, obtaining a different exponent estimate based on (\ref{gl:etf-plateau}) 
at least offers a possibility to control {\it a posteriori} previous estimates, since the systematic errors
should be different. If both scaling relations (\ref{gl:etf-plateau}) can be used, one could obtain simultaneously $\lambda$ and $\lambda/\mathpzc{z}$. 

The physical picture behind this discussion is based on the characteristic length scale $\ell(t)$. For the infinite system, one possible way of estimating it uses
the scaling of the second moment of $C(t;\vec{r})=C(t,t;\vec{r})$, namely  
\BD
\ell^2(t) =  \frac{\int_{\mathbb{R}^d}\!\D \vec{r}\: |\vec{r}|^2\: C(t;\vec{r})}{\int_{\mathbb{R}^d}\!\D \vec{r}\: C(t;\vec{r})} \sim t^{2/\mathpzc{z}}
\ED
which is characterised by the dynamical exponent. In a finite volume, 
one is then led to consider the second moment of the single-time correlator $C(t;\vec{r};N^{-1})$, using (\ref{25}) as follows
\BEA
\ell^2(t;N^{-1}) &=& 
\frac{\int_{\mathbb{R}^d} \!\D\vec{r}\: \vec{r}^2 C(t;\vec{r};1/N)}{\int_{\mathbb{R}^d} \!\D\vec{r}\: C(t;\vec{r};1/N)} 
\hspace{1.6cm}= \frac{\int_{\mathbb{R}^d} \!\D\vec{r}\: \vec{r}^2 s^{-b} \cdot 1 \cdot 
                \mathscr{F}_C\left({\vec{r}}{t^{-1/\mathpzc{z}}},{N}{t^{-1/\mathpzc{z}}}\right)}{\int_{\mathbb{R}^d} \!\D\vec{r}\: s^{-b} \cdot 1 \cdot 
                \mathscr{F}_C\left({\vec{r}}{t^{-1/\mathpzc{z}}},{N}{t^{-1/\mathpzc{z}}}\right)} 
\nonumber \\
&=& t^{2/\mathpzc{z}}\: \frac{\int_{\mathbb{R}^d} \!\D\vec{u}\: \vec{u}^2 
              \mathscr{F}_C\left(\vec{u},{N}{t^{-1/\mathpzc{z}}}\right)}{\int_{\mathbb{R}^d} \!\D\vec{u}\: 
              \mathscr{F}_C\left(\vec{u},{N}{t^{-1/\mathpzc{z}}}\right)}
\hspace{0.1cm}= t^{2/\mathpzc{z}} \mathscr{F}_{\ell}\left(\frac{N}{t^{1/\mathpzc{z}}}\right)
\label{3.16}
\EEA
where the finite-size scaling function $\mathscr{F}_{\ell}(v)$ remains after the integrations over $\vec{u}$ have been carried out. Now, if $N t^{-1/\mathpzc{z}}\gg 1$, the
limit $\mathscr{F}_{\ell}(\infty)=\mbox{\rm cste.}$ and if $N t^{-1/\mathpzc{z}}\ll 1$, the asymptotics $\mathscr{F}_{\ell}(v)\stackrel{v\ll 1}\sim v^2$ 
describe together the cross-over from the infinite-system growth to the saturation regime when the effects of the finite volume become tangible. 
This finite-size cross-over has been routinely observed in numerical simulations of lattice systems, 
e.g. \cite{Chris17,Chris19,Chris20,Chris21,Janke23} for recent examples, and the
finite-size scaling of the length scale $\ell(t)$ in (\ref{3.16}) 
may be viewed as a further corollary of the dynamical finite-size scaling (\ref{25}). 

Finally, if $d^*<d$, the system is no longer fully finite and one might expect a cross-over to an effective system in $d-d^*$ dimensions.  
The example of the spherical model (see appendix~D) makes it clear that there is no longer a plateau. More results
in different models will be needed to build up an intuition on the behaviour of the finite-size scaling function $\mathscr{F}_C(u)$.

\subsection{Finite-size effects in responses}  \label{sec:3.4}

Similarly, the finite-size scaling of auto-response functions can be analysed. 
Considering the same hyper-cubic geometry as before, we can repeat the formal calculations and find
\BEQ \label{gl:Rfss}
R\left(t,s;\vec{0};\frac{1}{N}\right) = s^{-1-a} \left(\frac{t}{s}\right)^{-\lambda/\mathpzc{z}} \mathscr{F}_R\left( \frac{N}{t^{1/\mathpzc{z}}}\right)
\EEQ
The behaviour of the system follows from the still largely unknown properties of the finite-size scaling function $\mathscr{F}_R(u)$.  
As before, we expect $\mathscr{F}_R(u)$ to depend also on the boundary conditions.  
It is tempting to {\em assume}, and by analogy with the auto-correlator, that for a fully finite system, the two-time response would converge to a plateau
whose height should be independent of the observation time $t$ (or equivalently $y=t/s$). 
If that should indeed turn out to be the case, then one may proceed as for the auto-correlator treated before. For the finite-size scaling function $\mathscr{F}_R$ one
will assume that it obeys $\mathscr{F}_R(\infty)=\mbox{\rm cste.}$ and $\mathscr{F}_R(u)\stackrel{u\ll 1}{\sim} u^{-\lambda}$. 

\noindent
{\bf Proposition 4.} {\it For a fully finite ageing system quenched to $T\leq T_c$, and if $s\ll N^{\mathpzc{z}}$, 
and if the auto-response function converges 
to a plateau with height $R_{\infty}^{(2)}=\lim_{y\to\infty}  R(ys,s;\vec{0};1/N)$, this height scales as 
\BEA
\mbox{\it if $s$ is kept fixed, then} && R_{\infty}^{(2)} \sim N^{-\lambda} \nonumber \\[0.1cm]
\mbox{\it if $N$ is kept fixed, then} && R_{\infty}^{(2)} \sim s^{\lambda/\mathpzc{z} -a-1} 
\label{gl:plateauR}
\EEA
where the exponent $a$ is given by (\ref{gl:val-a}).} 

In appendix~D, we show that in the fully finite kinetic spherical model in $2<d<4$ dimensions and quenched to $T<T_c$, 
such a convergence to a plateau indeed occurs.
The two-time auto-response function indeed obeys the finite-size scaling (\ref{gl:Rfss}) and the plateau height obeys 
(\ref{gl:plateauR}) such that $\mathpzc{z}=2$ and
$\lambda=d/2$ are read off. In the fully connected $p=2$ spherical spin glass quenched to $T=0$, the exact solution \cite{Fydo15} for large times
$t\gtrsim t_{\rm cross}\sim N^{2/3}$ (and $s\ll N^{2/3}$) shows that the noise-averaged response
is indeed consistent with the finite-size scaling (\ref{gl:Rfss}) and the exponents $\mathpzc{z}=\frac{2}{3}$ 
and $\lambda=\demi$ as determined in the previous sub-section from the auto-correlator. 
We point out that the values of the exponents $a$ and $\lambda/\mathpzc{z}$ remain the {\em same} 
as for the infinite-size system, see appendix~D for more details. 
There is no plateau is this case. 

It is too early to try to recognise any pattern in the behaviour of $\mathscr{F}_R(u)$ from the single spherical model for $d<d^*$.

In addition, time-integrated response functions can be considered. 
To illustrate the idea, consider the `intermediate' integrated response \cite{Corb03a,Henk03PP,Henk04PP} 
\BEQ
\chi_{\rm int}\left(t,s;\frac{1}{N}\right) := \int_{s/2}^s \!\D u\: R\left(t,u;\vec{0};\frac{1}{N}\right)
\EEQ
because more conventional integrated responses such as the thermoremanent magnetisation ({\sc trm}) 
or the zero-field-cooled susceptibility ({\sc zfc}) may be affected by strong corrections to
the scaling behaviour of interested. If $t\gg s$, we can use the factorised form (\ref{gl:Rfss}) 
of finite-size scaling where the finite-size scaling function merely
depends on the larger time $t$, and we find
\BEQ
\chi_{\rm int}\left(t,s;\frac{1}{N}\right) = s^{-a} \left(\frac{t}{s}\right)^{-\lambda/\mathpzc{z}} 
\underbrace{~\int_{1/2}^1 \!\D v\: v^{\lambda/\mathpzc{z}-1-a}~}_{\mbox{\rm\footnotesize cste.}}\:
\mathscr{F}_R\left(\frac{N}{t^{1/\mathpzc{z}}}\right)
\EEQ
where the finite-size scaling function $\mathscr{F}_R$ is the same as for the auto-response function. This gives

\noindent
{\bf Corollary 3.} {\it If for an ageing system quenched to $T\leq T_c$, with $s\ll N^{\mathpzc{z}}$, there is a plateau in the intermediate susceptibility defined as
$\chi_{\infty}^{(2)}=\lim_{y\to\infty} \chi_{\rm int}(ys,s;\vec{0};1/N)$ then it scales as}  
\BEA
\mbox{\it if $s$ is kept fixed, then} && \chi_{\infty}^{(2)} \sim N^{-\lambda} \nonumber \\
\mbox{\it if $N$ is kept fixed, then} && \chi_{\infty}^{(2)} \sim s^{\lambda/\mathpzc{z} -a} 
\label{gl:plateauchi}
\EEA

See appendix~D for an example in the context of the spherical model. 
The same kind of idea can also be used for the thermoremanent susceptibility, since for $t\gg s$ we re-use the factorised form (\ref{gl:Rfss}) of the response
function, with the same reasons as before. We can then apply once more the techniques of appendix~C to deal with the integral. If we now define
$\chi_{\infty}^{(2)}=\lim_{y\to\infty} \chi_{\rm TRM}(ys,s;\vec{0};1/N)$, then we re-obtain once more the scaling (\ref{gl:plateauchi}).

Since response functions can be directly computed through certain correlation functions \cite{Chat03,Ricc03}, these formul{\ae} should provide
new routes for extracting estimates of $\lambda$, $\lambda/\mathpzc{z}$ or $a$. 
We are not aware of any further test of (\ref{gl:plateauR}) or (\ref{gl:plateauchi}) in another specific model. 

It will have to be seen in future work which representations of Local Scale-Invariance \cite{Henk94,Henk10} will be needed for the computation of the form of 
the finite-size scaling function $\mathscr{F}_R(u)$.

\subsection{Magnetic initial states}

\begin{figure}[t]
\includegraphics[width=.45\hsize]{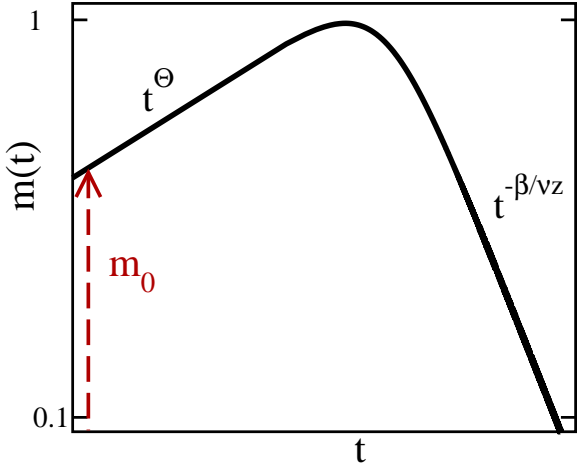} ~~~
\caption[fig4]{Time-dependence of the non-equilibrium magnetisation $m(t)$. \label{fig4}}
\end{figure}

For a quench onto criticality $T=T_c$, with a {\em non-}conserved order-parameter (model A), 
the magnetisation $m(t)$ has a non-trivial time-evolution \cite{Jans89} sketched in figure~\ref{fig4}. If the system
is prepared with a small initial magnetisation $\bigl|m(0)\bigr|=|m_0|\ll 1$ but is otherwise uncorrelated, 
there exists a short-time critical regime where $m(t)\sim t^{\Theta}$.
The {\em slip exponent} $\Theta$ is independent\footnote{This holds for model-A dynamics \cite{Jans89} 
as well as for model-C dynamics \cite{Oerd93,Cala03,Nand19,Nand20} (conserved energy density). 
For a conserved order-parameter (model B), clearly $\Theta=0$. 
For systems such as directed percolation (Reggeon field-theory/contact process), rapidity-reversal symmetry relates $\Theta$ (or $\lambda$) to stationary exponents, 
but the generic conclusions (\ref{JSS}) and (\ref{37}) remain valid \cite{Baum07b}.}  
of the equilibrium critical exponents and $\mathpzc{z}$.
Its value is found from deep field-theoretic considerations based
on the renormalisation group and short-time operator product expansions \cite{Jans89,Jans92}. 
One has the celebrated {\sc Janssen-Schaub-Schmittmann} ({\sc jss}) scaling relation \cite{Jans89}
\BEQ \label{JSS}
\Theta = \frac{d-\lambda}{\mathpzc{z}}
\EEQ
We shall meet this important scaling relation (originally established only for critical systems) again below, in different physical contexts. 
Returning to figure~\ref{fig4}, after passing through a maximum, the
magnetisation crosses over into a regime of critical decay, according to $m(t)\sim t^{-\beta/(\nu\mathpzc{z})}$, 
where $\beta$ and $\nu$ are standard equilibrium critical exponents \cite{Alba11,Taeu14}. 
Can one understand this behaviour and the existence of two distinct scaling regimes in figure~\ref{fig4}, 
in terms of a scaling description based on (\ref{gl:hyp})~? 
%
%

To answer this, we shall first relate the behaviour of $m(t)$ to a (global) response function. In order to see how this comes about, 
recall from Janssen-de Dominicis non-equilibrium field-theory \cite{Domi76,Jans76,Jans92,Taeu14} the
calculation of averages of an observable $A$
\BEQ \label{dynft}
\bigl\langle A\bigr\rangle = \int \mathscr{D}\phi\mathscr{D}\wit{\phi}\; A[\phi]\, e^{-{\cal J}[\phi,\wit{\phi}]}
\EEQ
where the action ${\cal J}[\phi,\wit{\phi}]={\cal J}_b[\phi,\wit{\phi}]+{\cal J}_{\rm ini}[\wit{\phi}]$ is decomposed into a bulk term and an initial term
(the explicit form (\ref{dynact}) is for non-conversed model-A dynamics)  
\begin{subequations} \label{dynact}
\begin{align}
{\cal J}_b[\phi,\wit{\phi}] &= \int \!\D t\D\vec{r}\: \left( \wit{\phi} \left( \partial_t - \Delta_{\vec{r}} - V'[\phi]\right)\phi - T \wit{\phi}^2 \right) \\
{\cal J}_{\rm ini}[\wit{\phi}] &= \int \!\D\vec{r}\: \left( \frac{1}{2\tau_0} \wit{\phi}^2(0,\vec{r}) - m_0 \wit{\phi}(0,\vec{r}) \right) \label{dynact-ini}
\end{align}
\end{subequations}
(with the usual re-scalings) which describes in the bulk a thermal white noise of temperature $T$, 
and an initially magnetised state of average magnetisation $m_0$ together with gaussian
fluctuations of width $\tau_0$. This implies the non-trivial short-time relationship $\phi(0,\vec{r})=\frac{1}{\tau_0}\wit{\phi}(0,\vec{r})$ \cite{Jans92}. 
Herein, `initial time' means of course a time-scale $\tau_{\rm mic}$ at the beginning of the scaling regime. 
The bulk action at temperature $T=0$, ${\cal J}_0[\phi,\wit{\phi}]  = \lim_{T\to 0} {\cal J}_b[\phi,\wit{\phi}]$ 
is called the {\em deterministic action} which does not contain any noise contribution. 
In analogy with (\ref{dynft}), {\em deterministic averages} $\bigl\langle \cdot \bigr\rangle_0$ can be defined as
\BEQ
\bigl\langle A\bigr\rangle_0 = \int \mathscr{D}\phi\mathscr{D}\wit{\phi}\; A[\phi]\, e^{-{\cal J}_0[\phi,\wit{\phi}]}
\EEQ
where in (\ref{dynft}) the action ${\cal J}$ is replaced by its deterministic part ${\cal J}_0$. 
Either from causality considerations \cite{Jans89,Jans92,Taeu14} or Local-Scale-Invariance ({\sc lsi}) \cite{Pico04} one obtains the Bargman su\-per\-se\-lec\-tion rules
\BEQ \label{Bargman}
\left\langle \overbrace{~\phi \cdots \phi~}^{\mbox{\rm ~~$n$ times~~}} 
             \overbrace{ ~\wit{\phi} \cdots \wit{\phi}~}^{\mbox{\rm ~~$m$ times~~}}\right\rangle_0 \sim \delta_{n,m}
\EEQ
for the deterministic averages. This means that only observables built from an equal number of order-parameters $\phi$ 
and conjugate response operators $\wit{\phi}$ can have non-vanishing deterministic averages. 

Full noisy averages should now be computed from (\ref{dynft}). They can be reduced \cite{Pico04} to deterministic averages by formally expanding in powers of 
$T,\frac{1}{\tau_0}$ and $m_0$ to all orders. It follows that the time-dependent magnetisation can be obtained as
\BEQ \label{gl:m0R}
m(t) = \bigl\langle \phi(t,\vec{0})\bigr\rangle 
= m_0 \int_{\mathbb{R}^d} \!\!\D\vec{r}\: \bigl\langle \phi(t,\vec{0}) \wit{\phi}(0,\vec{r}) \bigr\rangle_0 = m_0 \int_{\mathbb{R}^d} \!\!\D\vec{r}\: R(t,0;\vec{r}) 
= m_0 \wht{R}(t,0;\vec{0})
\EEQ
This follows from (\ref{dynft}) and the form of the action ${\cal J}[\phi,\wit{\phi}]$ detailed in (\ref{dynact}). 
Namely, the only term admissible, because of the Bargman rule
(\ref{Bargman}), is the one linear in $m_0$ as can be seen from (\ref{dynact-ini}). 
By definition, this is a response function and in the last step, we introduced the spatial Fourier transform
\BEQ \label{Fourier} 
\wht{R}(t,s;\vec{q}) = \int_{\mathbb{R}^d} \!\D\vec{r}\; e^{-\II \vec{q}\cdot\vec{r}} R(t,s;\vec{r})
\EEQ
Eq.~(\ref{gl:m0R}) is the starting point of our scaling analysis. 
The formally vanishing waiting time $s=0$ in (\ref{gl:m0R}) should be physically interpreted as a microscopically small
time $s_{\rm mic}\ll t$ at the beginning of the scaling regime. 

In the physical situation at hand, the initial magnetisation $m_0$ should be considered as a further dimensionful scaling variable, 
with scaling dimension $x_0$. The co-variance conditions of the global response function $\wht{R} = \wht{R}(t,s;\vec{0};m_0)$ then read
\begin{subequations} \label{gl:m0}
\begin{align}
X_{-1} \wht{R} &= \left( -\partial_t - \partial_s + \frac{\xi}{t} + \frac{\wit{\xi}}{s} \right) \wht{R} =  0 \label{gl:m0-1} \\
X_{0}  \wht{R} &= \left( -t\partial_t - s\partial_s - \frac{x_0}{\mathpzc{z}} m_0\partial_{m_0} -2\delta +\frac{d}{\mathpzc{z}} +\xi +\wit{\xi}\right) \wht{R} = 0 
\label{gl:m00}
\end{align}
\end{subequations}
In analogy to what was done before, solving (\ref{gl:m0}) gives for $t/s\gg 1$ the asymptotics
\BEQ
\wht{R}(t,s;\vec{0};m_0) = 
s^{-2\delta+d/\mathpzc{z}+\xi+\wit{\xi}} \left( \frac{t}{s}\right)^{-2\delta+d/\mathpzc{z}+\xi} \mathscr{F}_m\left(m_0 t^{-x_0/\mathpzc{z}}\right)
\EEQ
with the scaling function $\mathscr{F}_m(u)$ of the single argument $u=m_0 t^{-x_0/\mathpzc{z}}$. 
Since we need the response with respect to an `initial' perturbation, we must now send $s\to s_{\rm mic}$ and then can absorb it into the scaling function. 
Combination with (\ref{gl:m0R}) gives for the sought time-dependent magnetisation
\BEQ \label{36}
m(t) = m_0\: t^{\Theta} \mathscr{F}_m\left( m_0\: t^{-x_0/\mathpzc{z}}\right) \;\; , \;\;
\Theta = \frac{d}{\mathpzc{z}}+\xi-2\delta = \frac{d-\lambda}{\mathpzc{z}}
\EEQ
Herein, we identified the slip exponent and then observe that with (\ref{18a}) we indeed reproduce the {\sc jss} 
scaling relation (\ref{JSS}) which provides the sought-after relationship with field-theory \cite{Taeu14}. The scaling function $\mathscr{F}_m$ interpolates between
the two regimes of non-equilibrium critical scaling, through the assumed properties (i) $\mathscr{F}_m(0)=\mbox{\rm cste.}$ 
which reproduces the short-time scaling regime and (ii) $\mathscr{F}_m(u)\sim u^{-1}$ for $u\gg 1$ 
which makes the late-time scaling regime independent of $m_0$ and which also fixes $x_0$. \\

\noindent
{\bf Proposition 5:} {\it A critical system, with a non-conserved order-parameter, 
a small initial magnetisation $m_0$ but otherwise initially uncorrelated, has the time-dependent magnetisation
\BEQ \label{37}
m(t) = m_0\: t^{\Theta} \mathscr{F}_m\left( m_0\: t^{\Theta +\beta/(\nu\mathpzc{z})}\right) 
\EEQ
where $\Theta$ is given by (\ref{JSS}) and $\beta,\nu$ are standard equilibrium critical exponents.} 

Hence the well-known qualitative behaviour of figure~\ref{fig4} is reproduced. Although derived here for model A \cite{Jans89}, 
the end result (\ref{37}) holds for model C as well \cite{Oerd93,Cala03}. 
Given the generic relationship (\ref{gl:m0R}), the difficult proof \cite{Jans89} of the {\sc jss} scaling relation (\ref{JSS}) 
reduces here to a simple Fourier transform. Since the scaling function stems from a response
function, it is conceivable that larger dynamical symmetries (e.g. local scale-invariance) might be brought 
to furnish further details on its behaviour, where existing exact results of the spherical model \cite{Dutt08} may become useful.  We hope to return to this elsewhere. 

This result should only be applicable to critical quenches. Any attempt to prepare an initially magnetised state should lead for 
$T<T_c$ to a rapid relaxation towards one of the several distinct
equilibrium states in the phase-coexistence region such that the system will leave the regions in parameter space where dynamical scaling holds. 

\subsection{Global correlators and responses}

While all observables discussed so far have been {\em local} correlators or responses, 
we now consider {\em global} quantities, which are also often studied. For example, in an Ising model with spins $\sigma_n=\pm 1$ at 
each site $n$ of a lattice $\Lambda\subset\mathbb{Z}^d$, the {\em global} spin-spin correlator is defined as
\BEQ \label{gl:defQ}
Q(t,s) := \frac{1}{|\Lambda|} \left\langle \sum_{n,m\in\Lambda} \sigma_n(t) \sigma_m(s) \right\rangle
\EEQ
where $\bigl|\Lambda\bigr|$ is the number of sites of the lattice. 
Analogous definitions apply to other spin models. What can be said on the long-time behaviour of such observables~? \\

\noindent
{\bf Proposition 6:} {\it For a magnetic system quenched to $T\leq T_c$, 
the global spin-spin correlator (\ref{gl:defQ}) obeys for $t\gg s$ the scaling form
\begin{subequations}
\BEQ \label{eq:Qscal}
Q(t,s) = s^{-b+d/\mathpzc{z}} f_Q\left(\frac{t}{s}\right) \;\; ~~,~~ \;\; f_{Q}(y) \stackrel{y\gg 1}{\simeq} f_{\infty,Q}\, y^{\Theta}
\EEQ
with the constant $f_{\infty,Q}$ and the slip exponent $\Theta$ is given by (\ref{JSS}). In particular, for the correlation with the initial state  
\BEQ \label{eq:QTheta}
Q(t,0) \sim t^{\Theta} 
\EEQ
\end{subequations}}
In (\ref{eq:QTheta}), formally setting $s=0$ should be understood as a short-hand for the limit $s\to s_{\rm mic}$ to a microscopic reference time. 

This extends the validity of the scaling relation (\ref{JSS}) from the critical point, 
where it was originally derived \cite{Jans89}, to the entire low-temperature phase $T<T_c$. 

{\bf Proof:} At criticality, the result has been known since a long time \cite{Tome98}. It is derived here for all $T\leq T_c$ 
as a further example of generalised time-translation-invariance (\ref{gl:X-1gen}). 
Taking into account spatial translation-invariance, in the continuum limit the discrete sums over lattice sites 
are replaced by integrations over a spatial domain $V\subset \mathbb{R}^d$. 
Sending the integration volume $|V|$ to infinity, one has  
\BD
Q(t,s) = \lim_{|V|\to \infty}\frac{1}{|V|} \int_{V\times V} \!\D\vec{r}\D\vec{r}'\; \left\langle \phi(t,\vec{r}) \phi(s,\vec{r}')\right\rangle 
= \int_{\mathbb{R}^d} \!\D\vec{r}\; {C}(t,s;\vec{r})
= \wht{C}(t,s;\vec{0})
\ED
where the definition (\ref{Fourier}) of the Fourier transform was used. Recall the scaling forms (\ref{gl:2},\ref{gl:Cautoskal}) and carry out the
spatial integration. This gives for the $t\gg s$ asymptotics
\BEQ \label{40}
\wht{C}(t,s;\vec{0}) = s^{-b+\lambda/\mathpzc{z}} t^{-\lambda/\mathpzc{z}+d/\mathpzc{z}} 
                       \underbrace{ ~\int_{\mathbb{R}^d} \!\D\vec{u}\: \mathscr{F}_C(|\vec{u}|)~}_{\mbox{\rm\scriptsize = cste.}} 
\sim \left\{ \begin{array}{ll} s^{-\Theta+(2-\eta)/\mathpzc{z}}\, t^{\Theta} & \mbox{\rm ~~if $T=T_c$} \\[0.2cm]
                               s^{-\Theta+d/\mathpzc{z}}\, t^{\Theta}        & \mbox{\rm ~~if $T<T_c$}
             \end{array} \right.
\EEQ
where the identifications (\ref{18},\ref{36}) were used, 
which relate the slip exponent $\Theta$ with the autocorrelation exponent $\lambda$. This is (\ref{eq:Qscal}). 
Finally, the mathematical limit $s\to 0$ corresponds physically to the limit $s\to s_{\rm mic}$, 
where $s_{\rm mic}$ is a microscopic time-scale for the onset of the
dynamic scaling regime. Hence the dependence on $s_{\rm mic}$ can be absorbed into the scaling function 
$\mathscr{F}_C$. Then $Q(t,0)\sim Q(t,s_{\rm mic})\sim t^{\Theta}$ as asserted.
\hfill ~q.e.d.  

For non-equilibrium critical dynamics, eq.~(\ref{eq:QTheta}) 
has been routinely used to measure the exponent $\Theta$ at $T=T_c$ (or, via (\ref{JSS}), equivalently the autocorrelation exponent
$\lambda$, if $\mathpzc{z}$ is known) \cite{Jans89,Tome98,Silva23}. For systems quenched onto their critical point, the study of their universal short-time dynamics
has become a essential part of the tools of theoretical analysis, as reviewed in considerable detail in \cite{Zhen98,Alba11}. 
The derivation given here makes it clear that eq.~(\ref{eq:QTheta}) 
along with the scaling relation (\ref{JSS}) also furnishes an alternative method to find $\lambda$ from 
simulations for quenches into the low-temperature phase $T<T_c$. We are not aware yet of any numerical work along this line.

For estimating the dynamical exponent $\mathpzc{z}$, computing $\ell(t)$ from a second moment is tedious. 
An often-used alternative relies on the scaling of the single-time correlator, 
e.g. for phase-ordering where $b=0$ one may look for pairs $(t,\ell(t))$ by solving numerically a scaling equation such as 
$C(t,\vec{r})=F_C(|\vec{r}|/\ell(t))=0.5$ \cite{Maju17,Maju20}. Yet a further alternative
studies the short-time dynamics of the squared magnetisation $\bigl\langle m^2(t)\bigr\rangle$ \cite{Janke23}. 
Repeating the above arguments, notably eq.~(\ref{40}), it is easily seen that
\BEQ \label{41}
\bigl\langle m^2(t)\bigr\rangle \sim \wht{C}(t,t;\vec{0}) \sim t^{d/\mathpzc{z} - b} \sim 
\left\{ \begin{array}{ll} t^{d/\mathpzc{z}} & \mbox{\rm ~~if $T<T_c$} \\ t^{(2-\eta)/\mathpzc{z}} & \mbox{\rm ~~if $T=T_c$} \end{array} \right.
\EEQ
and in agreement with earlier heuristic arguments \cite{Huse89,Tome98,Silva02,Silva23,Janke23}. 
The numerical requirements of this technique are somewhat smaller than in the other methods mentioned above. 
In recent applications to phase-ordering kinetics in the $2D$ Ising model
it was concluded that as a tool for numerical computation the accuracy of (\ref{41}) is at least equal to the one of other techniques \cite{Janke23}. \\

\noindent
{\bf Proposition 7:} {\it In magnetic systems quenched to $T\leq T_c$, the time-dependent averaged squared magnetisation scales as
\BEQ
\bigl\langle m^2(t)\bigr\rangle \sim t^{d/\mathpzc{z} - b} = 
\left\{ \begin{array}{ll} t^{d/\mathpzc{z}}        & \mbox{\it ~~if $T<T_c$ {\rm ~~\cite{Janke23}}} \\[0.2cm] 
                          t^{(2-\eta)/\mathpzc{z}} & \mbox{\it ~~if $T=T_c$ {\rm ~~\cite{Huse89,Tome98}}} 
        \end{array} \right.
\EEQ
where $\eta$ is a standard equilibrium critical exponent.}\\

Numerous applications of these well-established results exist in the litt\'erature, see e.g. \cite{Zhen98,Alba11} for reviews.  

Similarly, we may also look at global response functions defined as 
\BEQ \label{gl:defP}
P(t,s) := \frac{1}{|\Lambda|} \left\langle \sum_{n,m\in\Lambda} R_{n,m}(t,s) \right\rangle
\EEQ
where $R_{n,m}(t,s)$ is the two-time response function between the lattice sites $n,m\in\Lambda$. 
In the continuum limit, we can write in analogy with what was done above
\BD
P(t,s) = \lim_{|V|\to \infty}\frac{1}{|V|} \int_{V\times V} \!\D\vec{r}\D\vec{r}'\; R(t,s;\vec{r}-\vec{r}') 
= \int_{\mathbb{R}^d} \!\D\vec{r}\; {R}(t,s;\vec{r})
= \wht{R}(t,s;\vec{0})
\ED
with the definition (\ref{Fourier}) of the Fourier transform. Using the scaling relation (\ref{gl:val-a}) 
of the local response, we recall the distinction into classes L and S and find 
\BEQ 
\wht{R}(t,s;\vec{0}) = s^{-1-a+\lambda/\mathpzc{z}} t^{-\lambda/\mathpzc{z}+d/\mathpzc{z}} 
                       \underbrace{ ~\int_{\mathbb{R}^d} \!\D\vec{u}\: \mathscr{F}_R(|\vec{u}|)~}_{\mbox{\rm\scriptsize = cste.}} 
\sim \left\{ \begin{array}{ll} s^{-\Theta-1+(2-\eta)/\mathpzc{z}}\, t^{\Theta} & \mbox{\rm ~~class L} \\[0.2cm]
                               s^{-\Theta-1+(d-1)/\mathpzc{z}}\, t^{\Theta}    & \mbox{\rm ~~class S}
             \end{array} \right.
\EEQ
We summarise this as follows, with the same notations and conventions as before.   

\noindent
{\bf Corollary 4:} {\it For a magnetic system quenched to $T\leq T_c$, 
the global spin response function (\ref{gl:defP}) obeys for $t\gg s$ the scaling form
\begin{subequations}
\BEQ \label{eq:Pscal}
P(t,s) = s^{-1-a+d/\mathpzc{z}} f_P\left(\frac{t}{s}\right) \;\; ~~,~~ \;\; f_{P}(y) \stackrel{y\gg 1}{\simeq} f_{\infty,P}\, y^{\Theta}
\EEQ
with the constant $f_{\infty,P}$. For the global response with respect to the initial state
\BEQ \label{eq:PTheta}
P(t,0) \sim t^{\Theta} 
\EEQ
\end{subequations}} 

At the critical point, eqs.~(\ref{eq:Qscal},\ref{eq:Pscal}) are known from a detailed and deep study in non-equilibrium field theory \cite[eqs. (86,87)]{Cala05} and
these apply for dynamics of models A (including tri-critical points, and weakly diluted Ising models with uncorrelated impurities), C, E and G. 
The conserved model-B dynamics has of course $\Theta=0$. Global correlators and responses were studied, for a non-conserved order-parameter, 
in the spherical model quenched onto $T=T_c$ \cite{Anni06,Anni08} and in the $1D$ Glauber-Ising model quenched to $T=0$ \cite{Maye04}, respectively. 


\subsection{Finite-size effects in global observables}

When we consider global correlators in the finite-size hyper-cubic geometry studied before, we can follow the lines of previous discussions. 
The finite-size global correlator $\wht{C}(t,s;\vec{0};N^{-1})$ must obey 
\begin{subequations} \label{gl:Q0}
\begin{align}
X_{-1} \wht{C} &= \left( -\partial_t - \partial_s + \frac{\xi}{t} + \frac{\wit{\xi}}{s} \right) \wht{C} =  0 \label{gl:Q0-1} \\
X_{0}  \wht{C} &= \left( -t\partial_t - s\partial_s + \frac{1}{\mathpzc{z}} \frac{1}{N}\partial_{1/N} -2(\delta-\xi) +\frac{d}{\mathpzc{z}}\right) \wht{C} = 0 
\label{gl:Q00}
\end{align}
\end{subequations}
which leads to, again with (\ref{JSS})  
\BEQ \label{3.42}
Q(t,s;N^{-1}) = \wht{C}(t,s;\vec{0};N^{-1}) = s^{-b+d/\mathpzc{z}} \left(\frac{t}{s}\right)^{\Theta} \wht{\mathscr{F}}_C\left(\frac{N}{t^{1/\mathpzc{z}}}\right)
\EEQ
with the finite-size scaling function $\wht{\mathscr{F}}_C$, which must satisfy $\wht{\mathscr{F}}_C(\infty)=\mbox{\rm cste.}$ 
and $\wht{\mathscr{F}}_C(u)\stackrel{u\ll 1}{\sim} u^{\mathpzc{z}\Theta}$. As before, we expect to find in the regime 
$\ell(t)\approx N$ and $\ell(s)\ll N$ a plateau with height
\BD
Q_{\infty}^{(2)} = \lim_{y\to \infty} Q(ys,s;N^{-1})
\ED 
Then, from (\ref{3.42}) 

\noindent
{\bf Corollary 5:} {\it For a magnetic system quenched to $T\leq T_c$, and if $\ell(t)\approx N$ and $\ell(s)\ll N$, the global spin-spin correlator $Q(t,s)$ 
converges to a plateau whose height scales as
\begin{subequations} \label{gl:plateauQ}
\begin{align}
\mbox{\it if the waiting time $s$ is fixed:~~~}  &  ~Q_{\infty}^{(2)} \sim N^{\mathpzc{z}\Theta}    = N^{d-\lambda}  \label{gl:plateauQN} \\
\mbox{\it if the system size  $N$  is fixed:~~~} &  ~Q_{\infty}^{(2)} \sim s^{\lambda/\mathpzc{z}-b} 
\end{align}
\end{subequations}
and where the use of (\ref{gl:val-b}) is understood, to distinguish between $T<T_c$ and $T=T_c$.}

\noindent 
Clearly, the scaling relation (\ref{JSS}) is available for all $T\leq T_c$ to relate the slip exponent $\Theta$ to the auto-correlator exponent $\lambda$. 

Similarly, for the global response function $\wht{R}(t,s;\vec{0};N^{-1})$ we must have  
\begin{subequations} \label{gl:P0}
\begin{align}
X_{-1} \wht{R} &= \left( -\partial_t - \partial_s + \frac{\xi}{t} + \frac{\wit{\xi}}{s} \right) \wht{R} =  0 \label{gl:P0-1} \\
X_{0}  \wht{R} &= \left( -t\partial_t - s\partial_s + \frac{1}{\mathpzc{z}} \frac{1}{N}\partial_{1/N} -2\delta +\frac{d}{\mathpzc{z}} +\xi +\wit{\xi}\right) \wht{R} = 0 
\label{gl:P00}
\end{align}
\end{subequations}
which gives, with the finite-size scaling function $\wht{\mathscr{F}}_R$,
\BEQ
P(t,s;N^{-1}) = \wht{R}(t,s;\vec{0};N^{-1}) = s^{-1-a+d/\mathpzc{z}} \left(\frac{t}{s}\right)^{\Theta} \wht{\mathscr{F}}_R\left(\frac{N}{t^{1/\mathpzc{z}}}\right)
\EEQ
and the analogous asymptotic properties as before. 

\noindent
{\bf Corollary 6:} {\it For a magnetic system quenched to $T\leq T_c$, and if $\ell(t)\approx N$ and $\ell(s)\ll N$, and if the global spin response function $P(t,s)$ 
converges to a plateau $P_{\infty}^{(2)}=\lim_{y\to\infty} P(ys,s;N^{-1})$, its height scales as
\begin{subequations}  \label{gl:plateauP}
\begin{align}
\mbox{\it if the waiting time $s$ is fixed:~~~}  &  ~P_{\infty}^{(2)} \sim N^{\mathpzc{z}\Theta}      = N^{d-\lambda}   \label{gl:plateauPN} \\
\mbox{\it if the system size  $N$  is fixed:~~~} &  ~P_{\infty}^{(2)} \sim s^{\lambda/\mathpzc{z}-1-a} 
\end{align}
\end{subequations}
and where the values (\ref{gl:val-a}) must be used to distinguish between the classes L and S.}

The only application of this we are aware of occurs in the $1D$ kinetic Ising model with Kimball-Deker-Haake ({\sc kdh}) dynamics 
\cite{Kimb79,Deke79,Dutt09}. This is a model
of non-conserved single spin-flip dynamics, but because of its unusual rates, the dynamical exponent $\mathpzc{z}=4$ \cite{Kimb79,Deke79}. 
For a fully disordered initial state, on a periodic chain of $N$ sites, and after a quench to $T=0$, it can be
shown exactly that for large times $Q(t,0;N^{-1})=\frac{1}{3}+{\rm O}(e^{-t})$. Via (\ref{gl:plateauQN}), this can be compared to the plateau value 
$Q_{\infty}^{(2)}\sim N^{\mathpzc{z}\Theta}=N^{1-\lambda}$. We read off
$\lambda=1$ \cite{Dutt09}.\footnote{The same conclusion $\lambda=1$ is reached from calculating 
$C(t,0)=N^{-1}+{\rm O}\bigl(e^{-t}\bigr)$ and (\ref{gl:etf-plateauN}) \cite{Dutt09}.} 
This is in agreement with the global response with respect to the initial state in the infinite-size system, for which 
$P(t,0)=\wht{R}(t,0)=\big(1-\tanh \frac{1}{T_{\rm init}}\bigr)^3 +{\rm O}(e^{-t})$ \cite{Dutt09} where $T_{\rm init}$ is the temperature of the initial state. 
Comparison with (\ref{eq:PTheta}) leads back to $\lambda=1$ so that we have also re-confirmed
the expected equality between auto-correlation and auto-response exponents.

The main point of this long section has been that all results derived here, 
which together constitute the basic phenomenology of \BLAU{classical} non-equilibrium ageing, 
and almost all of them having been well-known (individually) since a long time, 
were seen to  have a single common origin, namely the generalised time-translation-invariance (\ref{gl:inv-1}) together
dynamical scaling (\ref{gl:inv0}), the latter adapted to the physical situation at hand. 
Besides, we merely used simple consistency constraints on the
various scaling functions which arose. Probably the main news is the extension of
the {\sc jss} scaling relation (\ref{JSS}) to all temperatures $T\leq T_c$; and the new finite-size scaling laws 
(\ref{gl:etf-plateau},\ref{gl:plateauR},\ref{gl:plateauchi},\ref{gl:plateauQ},\ref{gl:plateauP}) for the plateaux in fully finite systems. 

\section{The equation of motion} \label{sec:4}

\subsection{Phase-ordering kinetics: the basic idea} \label{sec:4.1}

We now turn to an analysis of the equation of motion for the order-parameter. We begin with the case of phase-ordering, where for a non-conserved model-A dynamics,
after a quench to $0<T<T_c$, one expects an equation of the form \cite{Bray94a,Maze06}\footnote{The underlying assumption of the applicability of the continuum limit
description is not valid in the $2D$ Ising model with long-ranged interaction, quenched to exactly $T=0$, where $z=\frac{4}{3}$ is found \cite{Agra21,Chris21}.}
\BEQ \label{gl:4.1}
\left( \partial_t - \frac{1}{2{\cal M}} \Delta_{\vec{r}} \right) \phi(t,\vec{r}) = - \frac{\delta {\cal V}[\phi]}{\delta \phi(t,\vec{r})} +\eta(t,\vec{r})
\EEQ
where ${\cal V}[\phi]$ is the coarse-grained functional for the interaction, $\Delta_{\vec{r}}$ is the spatial laplacian, 
$\eta(t,\vec{r})$ represents the thermal noise and ${\cal M}$ is a non-universal constant.  In principle, this equation must be solved in order to find correlators such as 
$\langle \phi(t,\vec{r})\phi(s,\vec{0})\rangle$. For this, the precise form of the interaction potential 
${\cal V}[\phi]$ is not really important, provided only that is possesses
at least two equally deep minima which model the distinct but equivalent macroscopic physical states, see also figure~\ref{fig1}. 
For short-ranged interactions, the thermal noise is irrelevant
in the renormalisation-group sense and can be dropped in an analysis of the leading scaling behaviour. On the other hand, a noisy initial state will have to be considered,
although one will usually admit that the averaged order-parameter vanishes: $\langle\phi\rangle=0$. We add that only initially un-correlated states are considered. 
Then a stability analysis of (\ref{gl:4.1}), see also figure~\ref{fig1}, 
shows that the initial state will be unstable and the system will rapidly evolve towards a spatially inhomogeneous state \cite{Bray94a}. 
This state is no longer described by (\ref{gl:4.1}) for the effective
long-time behaviour but we shall consider the more simple form (directly restricted to non-conserved model-A type dynamics) 
\BEQ \label{gl:4.2}
\left( \partial_t - \frac{1}{2{\cal M}} \Delta_{\vec{r}} \right) \phi(t,\vec{r}) = \phi^3(t,\vec{r}) 
\EEQ
This form is suggestive for the following reasons:
\begin{enumerate}
\item a term linear in $\phi(t,\vec{r})$ on the right-hand-side of (\ref{gl:4.2}) would break dynamical scaling
\item a term quadratic in $\phi(t,\vec{r})$ would break the global spin-reversal-invariance
\item a term cubic in $\phi(t,\vec{r})$ is the lowest-order term which may appear (we shall see below that higher-order terms will lead to corrections to scaling)
\item thermal noise will merely lead to corrections to scaling, hence it is left out
\item the dynamical exponent $\mathpzc{z}=2$ \cite{Bray94b} of systems with short-ranged interactions is included
\end{enumerate}

On the left-hand-side of (\ref{gl:4.2}) we have the Schr\"odinger operator $\mathscr{S}=\partial_t - \frac{1}{2{\cal M}}\Delta_{\vec{r}}$. 
Since according to our postulate (\ref{gl:hyp}), the dynamics far from equilibrium must be described by a modified representation and characterised by the constant
$\xi$, we consider the Schr\"odinger operator in the new representation and have
\BEQ \label{gl:4.3} 
\overline{\mathscr{S}} = e^{\xi \ln t} \mathscr{S} e^{-\xi \ln t} = \partial_t - \frac{\xi}{t} - \frac{1}{2{\cal M}} \Delta_{\vec{r}} 
\EEQ
which contains an additional $1/t$-potential and will act on the transformed order-parameter $\overline{\phi} = e^{\xi\ln t}\phi = t^{\xi}\phi$. 
The original equation (\ref{gl:4.2}) $\mathscr{S} \phi = \phi^3$ then becomes in the new basis
\BEQ \label{gl:4.4}
\left( t^{\xi}\, \mathscr{S}\, t^{-\xi} \right) \left( t^{\xi}\, \phi \right) = t^{\xi} \left( t^{-\xi}\, \overline{\phi}\, \right)^{3} 
~~\Longrightarrow~~ 
\overline{\mathscr{S}}\,\overline{\phi} = \left( \partial_t - \frac{\xi}{t} - \frac{1}{2{\cal M}} \Delta_{\vec{r}} \right) \overline{\phi} = t^{-2\xi}\, \overline{\phi}^{\,3}
\EEQ
and we recall that for phase-ordering kinetics $\delta_{\rm eff}=\delta-\xi=0$ such that $\overline{\phi}$ is dimensionless. 
Examining the long-time behaviour of (\ref{gl:4.4}), we must take into account the explicit $t$-dependence. We see that the 
$1/t$-potential will for large times dominate over against the non-linear term, provided that
\BEQ \label{gl:4.5}
2\xi > 1 ~~\Longleftrightarrow~~ \lambda > 1
\EEQ
where we recall (\ref{18b}).  In $d>2$ dimensions, this condition is always satisfied because of the {\sc yrd}-bound (\ref{gl:lambda}) which gives 
$\lambda \geq d/2 >1$. For $d=2$ dimensions, one has typically
$\lambda \approx 1.25 >1$ (see \cite{Henk10} and refs. therein) such that the criterion (\ref{gl:4.5}) is also satisfied.  
Evidently, higher-order non-linearities in (\ref{gl:4.2},\ref{gl:4.4}) will decrease more fast as $t\to\infty$ when (\ref{gl:4.5}) holds.  
We summarise:  \\

\noindent
{\bf Proposition 8.} {\it For phase-ordering kinetics after a quench to $0<T<T_c$, in $d\geq 2$ dimensions, local interactions and non-conserved model-A dynamics such that 
$\mathpzc{z}=2$ and $\lambda>1$, the long-time dynamical symmetries of eq.~(\ref{gl:4.2}) are obtained 
as the symmetries of the linear equation $\overline{\mathscr{S}}\,\overline{\phi}=0$, with the transformed Schr\"odinger operator (\ref{gl:4.3}).} 

This is a central result of this work: given the validity of the criterion (\ref{gl:4.5}), 
the dynamical symmetries of a linear equation will govern the leading long-time behaviour of phase-ordering kinetics
although the values of $\lambda$ and $\mathpzc{z}$, which were used here
as parameters, must still be found from a detailed study of the full non-linear problem (\ref{gl:4.1}).  
 
\noindent
{\bf Example:} the obvious case study is the spherical model, see section~\ref{sec:2}. 
The quenched equation of motion at $T<T_c(d)$ is, for the leading part of the order-parameter (and $T_c(d)>0$ for $d>2$) 
\BD
\left( \partial_t - \frac{1}{2{\cal M}} \Delta_{\vec{r}} - \frac{\digamma}{2}\frac{1}{t} \right) \overline{\phi}(t,\vec{r}) = 0 \;\; , \;\;
\digamma = -\frac{d}{2}  \tag{\ref{2.3a}}
\ED
The value $\digamma$ in the time-dependent term comes from the asymptotic solution of the spherical constraint
$\bigl(2\pi\bigr)^{-d}\int_{\cal B} \!\D\vec{r}\: \overline{\phi}^2(t,\vec{r}) =1$. 
We recognise the Schr\"odinger operator $\overline{\mathscr{S}}$ of (\ref{gl:4.4}), 
wherein the spherical model Lagrange multiplier $\mathfrak{z}(t)=\frac{\digamma}{2} t^{-1}=\xi\, t^{-1}$ comes from
the change of representation (\ref{gl:hyp},\ref{gl:4.3}). The dynamical symmetries\footnote{\BLAU{In equilibrium quantum systems, the hamiltonian $H$ plays the r\^ole of the
Schr\"odinger operator $\overline{\mathscr{S}}$ and a symmetry Lie algebra generator $X$ should satisfy $[H,X]=\mathfrak{x} X$ in analogy with (\ref{gl:4.6}). When $H(t)$ becomes time-dependent, 
the required time-ordering must be included as well \cite{Mede20a,Mede20b}.}} of the Schr\"odinger operator $\overline{\mathscr{S}}$ follow from
the commutators with the generators (\ref{gl:inv})
\BEQ \label{gl:4.6} 
\left[ \overline{\mathscr{S}}, X_{-1} \right] = 0 \;\;, \;\; \left[ \overline{\mathscr{S}}, X_{0} \right] = -\overline{\mathscr{S}}
\EEQ
such that any solution of $\overline{\mathscr{S}}\,\overline{\phi}=0$ is mapped onto another solution \cite{Nied74}: the kinetic spherical model (\ref{2.3a})  obeys
the symmetries of generalised time-translation and scaling-invariance. 

With the identifications (\ref{3.5}) appropriate for a response function, the exact two-time response function of the spherical model is for 
$t>s$ ($R_{(0)}$ is a normalisation constant) \cite{Henk06}
\BEQ \label{gl:4.7}
R(t,s;r) = \left\langle \overline{\phi}(t,\vec{r}) {\overline{\wit{\phi}}}(s,\vec{0})\right\rangle 
= R_{(0)}\, s^{-1-a} \left( \frac{t}{s}\right)^{1+a'-\lambda/2} 
\left(\frac{t}{s} -1 \right)^{-1-a'} \exp\left[ -\frac{\cal M}{2} \frac{r^2}{t-s} \right] 
\EEQ
where the exponents $a,a',\lambda$ are identified as
\BEQ
\frac{\lambda}{2} = 2\delta -\xi \;\; , \;\; 1+a = 2\delta -\xi - \wit{\xi} \;\; , \;\; a'-a = \xi +\wit{\xi}
\EEQ
and $\cal M$ is a non-universal, dimensionful constant. 
Since in the spherical model, for $T<T_c(d)$, one has $a=a'=\frac{d}{2}-1$ and $\lambda=\frac{d}{2}$ \cite{Godr00b} it follows 
\BEQ
\delta = \frac{d}{4} \;\; , \;\; \xi = - \wit{\xi} = \frac{d}{2} - \frac{\lambda}{2} = \frac{d}{4}
\EEQ
in agreement with the prediction $\delta=\xi$ for phase-ordering (section~\ref{sec:3}), as it should be. 

It is well-known that potentials with a $1/t$-form in $\overline{\mathscr{S}}$ 
arise in certain analytical studies on phase-ordering kinetics \cite{Oono88,Maze90,Maze04,Maze06}. 
They have also been found to be in good agreement with
recent experiments on phase-ordering in liquid crystals quenched into the ordered phase \cite{Alme21}.

\subsection{Phase-ordering kinetics: dimensionful coupling constants} \label{sec:4.2}

An immediate application of the above relationships to the phase-ordering kinetics of systems such as the $2D/3D$ Ising models is not possible. 

In order to understand the reason for this and to finally resolve this difficulty, we recall first that the Schr\"odinger operator $\mathscr{S}$ in (\ref{gl:4.3}) 
-- and we shall drop the bar on both $\mathscr{S}$ and $\phi,\wit{\phi}$ in this sub-section -- has a larger dynamical symmetry \cite{Nied74}, 
which is given by the generators of the Schr\"odinger Lie algebra $\langle X_{\pm 1,0}, Y_{\pm\demi},M_0\rangle$, see \cite{Nied72,Hage72,Duva24} and refs. therein.
The explicit form is \cite{Henk94,Henk06,Stoi22}, after application of (\ref{gl:hyp})
\begin{subequations}
\begin{align}
X_n &= -t^{n+1}\partial_t - \frac{n+1}{2}t^n r\partial_r -(n+1)\bigl(\delta-\xi\bigr) -n\xi t^n - \frac{n(n+1)}{4}{\cal M} t^{n-1} r^2 \label{gl:4.10a} \\
Y_m &= -t^{m+\demi}\partial_r - \bigl(m+\demi\bigr) {\cal M} t^{m-\demi} r \\
M_n &= -{\cal M} t^n 
\end{align}
\end{subequations}
These are indeed symmetries, only provided that $\delta=\frac{d}{4}$: then, because of the commutator relations 
\BEQ \label{gl:4.11}
\left[ {\mathscr{S}}, X_{-1} \right] = 0 \;\; , \;\;
\left[ {\mathscr{S}}, X_{0} \right] = -{\mathscr{S}} \;\; , \;\;
\left[ {\mathscr{S}}, X_{1} \right] = -2t{\mathscr{S}} - 2 \left( \delta - \frac{d}{4} \right)
\EEQ
(the commutators of the generators $Y_{\pm\demi}$, $M_0$ of the Schr\"odinger algebra with $\mathscr{S}$ vanish) 
imply that any solution of $\mathscr{S}\phi=0$ is mapped via $\phi\mapsto \bigl( 1 +\vep X\bigr) \phi$ onto another solution since 
$\mathscr{S}\bigl( X\phi\bigr)=0$ \cite{Nied74}. 
While the condition on $\delta$ certainly holds true for the spherical model, in the Ising model $\delta$ depends differently on $d$. 

To avoid this, consider rather semi-linear equations of the form \cite{Boye76}
\BEQ \label{gl:4.12}
\mathscr{S} \Phi = F\bigl(t,\vec{r},g,\Phi,\wit{\Phi}\bigr)
\EEQ
where we also introduce a dimensionful coupling $g$. Notice that we do admit a possible dependence on both the order-parameter $\Phi$ and the response field $\wit{\Phi}$. 
We look for dynamical time-space symmetries of (\ref{gl:4.12}) with a generator
\BEQ \label{gl:4.13}
X\Phi = \left( a \partial_t + b \partial_{\vec{r}} + d \partial_g + c \right) \Phi \;\; , \;\; 
\left[ \mathscr{S}, X \right] = \mu_1 \mathscr{S} + \mu_2 
\EEQ
where $a,b,c,d$ and $\mu_{1,2}$ are to be found as functions of $t,\vec{r},g$ such that solutions of (\ref{gl:4.12}) are mapped onto other solutions. It can be shown
that this leads to \cite{Boye76,Stoi05}
\BEQ \label{gl:4.14}
\left( a \partial_t + b \partial_{\vec{r}} + d \partial_g - c\,\Phi\partial_{\Phi} - \wit{c}\,\wit{\Phi}\partial_{\wit{\Phi}} +c +\mu_1 \right) F(t,\vec{r},g,\Phi,\wit{\Phi}) 
+ \mu_2 \Phi = 0
\EEQ
That $g$ is a dimensionful coupling, with scaling dimension $\psi$, is expressed through the following generalisation of the generator (\ref{gl:4.10a}) \cite{Stoi05}
\BD
X_n = -t^{n+1}\partial_t - \frac{n+1}{2}t^n r\partial_r -(n+1)\bigl(\delta-\xi\bigr)t^n -n\xi t^n 
      - (n+1) \psi t^n g\partial_g -\frac{n(n+1)}{4}{\cal M} t^{n-1} r^2 \tag{\ref{gl:4.10a}'} 
\ED
This does not change the Lie algebra commutators. 
Because of spatial translation-invariance, generated by $Y_{-\demi}$, 
we can translate to the frame where $\vec{r}=\vec{0}$. Also, 
we restrict our search to such non-linearities which do {\em not} depend on the response field $\wit{\Phi}$. 
Specifically, dynamical invariance of the equation (\ref{gl:4.12}) 
under the time-translation $X_{-1}$, dilatations $X_0$ and special Schr\"odinger transformations $X_1$ leads via (\ref{gl:4.11},\ref{gl:4.13},\ref{gl:4.14}) to
\BEA
\left( -\partial_t - \frac{\xi}{t} \Phi\partial_{\Phi} + \frac{\xi}{t} \right) F(t,g,\Phi) &=& 0 \nonumber \\
\biggl( -t\partial_t - \psi g\partial_g +\bigl(\delta-\xi\bigr)\Phi\partial_{\Phi} -\delta +\xi -1 \biggr) F(t,g,\Phi) &=& 0 \\
\biggl( -t^2\partial_t -2\psi tg\partial_g +\bigl(2\delta-\xi\bigr)t\Phi\partial_{\Phi} -\bigl(2\delta-\xi+2\bigr)t\biggr) F(t,g,\Phi) &=& 2\left(\delta - \frac{d}{4}\right)\Phi 
\nonumber
\EEA
which can be simplified into
\BEA
\left( -\partial_t - \frac{\xi}{t} \Phi\partial_{\Phi} + \frac{\xi}{t} \right)       F(t,g,\Phi) &=& 0 \nonumber \\
\biggl( - \psi g\partial_g +\delta\,\Phi\partial_{\Phi} -\bigl(\delta+1\bigr) \biggr) F(t,g,\Phi) &=& 0 \\
\left( \delta - \frac{d}{4} \right) \Phi &=& 0 \nonumber
\EEA
The non-linearity then turns out to become
\BEQ
F(t,g,\Phi) = t^{-2\xi} g^{(2\delta+1)/\psi}\, \Phi^3 \,\mathfrak{F}\left( \Phi\, t^{-\xi}\, g^{\delta/\psi} \right) \;\; , \;\; \delta = \frac{d}{4}
\EEQ
where the function $\mathfrak{F}$ is still arbitrary. 
We shall take it to be a constant $\mathfrak{F}_0$ and then recover the equation of motion for the order-parameter, appropriate for phase-ordering kinetics
\BEQ \label{gl:4.18} 
\mathscr{S} \Phi = \left( \partial_t - \frac{\xi}{t} - \frac{1}{2{\cal M}} \Delta_{\vec{r}}\right) \Phi = \mathfrak{F}_0\: t^{-2\xi}\, g^{(2\delta+1)/\psi}\, \Phi^3 
= \mathfrak{F}_0\, g^{(d/2+1)/\psi}\: t^{-2\xi}\,\Phi^3
\EEQ
Phenomenologically, the dependence on $g$ will  be discarded. Satisfactorily, we have reproduced the equation of motion (\ref{gl:4.2}), 
which was argued for heuristically in the previous sub-section.

Having seen how to use a dimensionful coupling in the derivation of the effective equation of motion (\ref{gl:4.2},\ref{gl:4.18}) 
for large times, we now find the two-time response function. 
To do this, we factorise $\Phi= g \phi$ into a dimensionful coupling $g$ of scaling dimension $\psi$ and a field $\phi$ with the canonical dimension 
$\delta=\frac{d}{4}$. Analogously also write $\wit{\Phi}=\wit{g}\wit{\phi}$. The two-time response function then becomes
\BEQ \label{gl:4.19}
R(t,s;\vec{r}) = \left\langle \Phi(t,\vec{r})\wit{\Phi}(s,\vec{0})\right\rangle 
= g \wit{g} \left\langle \phi(t,\vec{r})\wit{\phi}(s,\vec{0})\right\rangle =: g\wit{g}\, \mathfrak{R}(t,s;\vec{r})
\EEQ
The function $\mathfrak{R}$ does not depend on $g,\wit{g}$. We are mainly interested in the auto-response and therefore set $\vec{r}=\vec{0}$. 
Since the non-linear term in (\ref{gl:4.18}) is irrelevant whenever the criterion (\ref{gl:4.5}) 
holds, we can use the Schr\"odinger-invariance of the generators $X_{\pm 1,0}$ of the remaining
linear equation. Then the co-variance conditions under $X_{\pm 1,0}$ turn into the following conditions for the auto-response 
\BEA
X_{-1} R(t,s) &=& \left( -\partial_t +\frac{\xi}{t} - \partial_s + \frac{\wit{\xi}}{s}\, \right)  g\wit{g}\, \mathfrak{R}(t,s) = 0 \nonumber \\
X_{0} R(t,s)  &=& \left( -t\partial_t -\bigl(\delta-\xi\bigr) - \psi -s\partial_s -\bigl({\delta} -\wit{\xi}\,\bigr) - {\psi} \right) g\wit{g}\, \mathfrak{R}(t,s) = 0 \\
X_{1} R(t,s)  &=& \left( -t^2\partial_t -\bigl(2\delta-\xi\bigr)t - 2\psi t -s^2\partial_s -\bigl(2{\delta} -\wit{\xi}\,\bigr)s - 2{\psi}s \right) g\wit{g}\, \mathfrak{R}(t,s) = 0 
\nonumber
\EEA
where we simply evaluated the explicit dependence on $g$ and $\wit{g}$ in (\ref{gl:4.19}). 
The solution of this system is a standard exercice \cite{Stoi05,Henk06,Henk10,Stoi22,Henk23b}. 
We shall not explicitly repeat this well-trodden path (it is enough to replace $\delta \mapsto \delta + \psi$) 
and just state that at the end, one is back to (\ref{gl:4.7}),
up to normalisations, but the exponents have now to be identified as follows  
\BEQ \label{gl:4.21}
a'-a = \xi + \wit{\xi} \;\; , \;\; \frac{\lambda}{2} = 2\delta +2\psi -\xi \;\; , \;\; 1+a = 2\delta +2\psi -\xi -\wit{\xi}
\EEQ
In addition, we have to set $\delta=\frac{d}{4}$. 
The space-dependent part of $R(t,s;\vec{r})$ would follow from the co-variance under $Y_{\pm\demi}$ and $M_0$ and is unchanged with respect to (\ref{gl:4.7}). 
The idea of \cite {Stoi05} means that in practise, we can forget about the constraint $\delta=\frac{d}{4}$ 
since through the introduction of a dimensionful coupling its effects can be absorbed into the
value of its scaling dimension $\psi$.

The generalised form (\ref{gl:4.7},\ref{gl:4.21}), or said differently: with $a,a',\lambda$ 
as free independent parameters, can be used in models different from the spherical model. 

\noindent
{\bf Proposition 9.} {\it For phase-ordering kinetics after a quench to $0<T<T_c$, for local interactions and non-conserved model-A dynamics such that 
$\mathpzc{z}=2$ and $\lambda>1$, the two-time response function is given by the form (\ref{gl:4.7}) and where the three exponents
$a,a',\lambda$ (and also the non-universal mass $\cal M$) can be chosen as free and independent parameters.}

\noindent
{\bf Example.} In the Glauber-Ising model in $d\geq 2$ dimensions and $0<T<T_c$, it is well-known that $a=\demi$ and $a'-a=0$ \cite{Henk01,Henk03}. 
The first and third condition (\ref{gl:4.21}) then give\footnote{The width of the interfaces in $2D/3D$ kinetic Ising models at $T<T_c$ scales as
$w^2(t) \sim t^{2\psi}$ \cite{Abra89}, with $\psi$ from (\ref{gl:4.22}). This should be compared to the linear size $\ell(t)^2\sim t$ of an ordered domain.} 
\BEQ \label{gl:4.22}
\xi + \wit{\xi} =0 \;\; , \;\; \frac{3}{2} = 2\delta +2\psi \;\; , \;\; \delta=\frac{d}{4} ~~\Longrightarrow~~ 2\psi = \frac{3-d}{2}
\EEQ
Consequently, we have with numerical approximations for $\lambda$, see \cite{Henk10} and refs. therein
\BEQ
\xi = 1+a -\frac{\lambda}{2} \simeq \left\{
\begin{array}{ll} \frac{3}{2} - \frac{5/4}{2} = \frac{7}{8}  > \frac{1}{2} & \mbox{\rm ~~;~ if $d=2$} \\[0.2cm]
                  \frac{3}{2} - \frac{1.6}{2} = \frac{7}{10} > \frac{1}{2} & \mbox{\rm ~~;~ if $d=3$}
\end{array} \right.
\EEQ
Hence the irrelevance criterion (\ref{gl:4.5}), of the non-linear term in eq.~(\ref{gl:4.18}), 
is indeed obeyed.\footnote{In quenches of the $3D$ Ising model to $T=0$, early simulations
on relatively small lattices apparently produced results different from the expected \cite{Bray94b} $z=2$. 
A detailed recent study undertakes to clarify the possible pre-asymptotic effects and still
expects $z=2$ to be recovered for truly large times \cite{Gess23}. 
Claims of a violation of the bound (\ref{gl:lambda}) and of a non-universal behaviour of $\lambda$ \cite{Vada19,Vada22} are carefully explained as arising from finite-size effects
and the value $\lambda=1.58(14)$ is quoted \cite{Gess24}, which obeys (\ref{gl:lambda}).}

Remarkably, there are long-standing confirmations, first of the auto-response \cite{Henk01}, second of the entire time-space response function \cite{Henk03}, 
in both the phase-ordering of the $2D$ as well as the $3D$ Glauber-Ising model, although both models are certainly not described by gaussian free fields. 
These tests are usually carried out using an integrated response, such as $\chi_{\rm TRM}(t,s)$ 
since trying to obtain $R(t,s)$ directly from Monte Carlo simulations will lead to very noisy data. 
Proposition~9 provides, \BLAU{for the first time,} a justification for this long-established practise. 

Analogous confirmations of the integrated prediction (\ref{gl:4.7}) of local-scale-invariance,  for quenches to $T<T_c$, 
have been obtained from simulations carried out for the $2D$ $q$-states Potts model \cite{Lorenz07a} ($q=2,3,8)$ and the $2D$ and $3D$ XY-model \cite{Abri03,Abri04}. 
A long list of tests of (\ref{gl:4.7}) is provided in \cite{Henk10}. In the known cases with $0<T<T_c$, one always finds $a'-a=\xi+\wit{\xi}=0$.

\subsection{Non-equilibrium critical dynamics} 

For a system quenched onto a critical point, the effective equation of motion for long times should take the form 
\BEQ \label{gl:4.24}
\left( \partial_t - \frac{1}{2{\cal M}} \Delta_{\vec{r}}^{\mathpzc{z}/2} \right) \phi(t,\vec{r}) = g \phi^3(t,\vec{r}) +\eta(t,\vec{r}) 
\EEQ
for a non-conserved model-A type dynamics. We modified the spatial part in order to take the non-trivial value of the dynamical exponent $\mathpzc{z}$ into account. 
More precisely, it is conceivable this term should be interpreted in terms of of some fractional derivative. 
As long as this commutes with the time $t$ and the time derivative $\partial_t$, 
we need not enter into the details of this since we shall concentrate on the local auto-response $R(t,s;\vec{0})$. 
On the right-hand-side, we merely retained the most relevant term of the critical point interactions \cite{Cala05,Taeu14}. 
The thermal noise $\eta(t,\vec{r})$ is now relevant and must be kept. 

We restrict our ambitions to a calculation of the two-time auto-response $R(t,s;\vec{0})$. Then any purely spatial parts in (\ref{gl:4.24}) can be dropped. As a response function, 
$R(t,s;\vec{0})$ does not depend explicitly on the noise such that we can restrict further to the {\em deterministic part} of (\ref{gl:4.24}), without the noise $\eta(t,\vec{r})$. 
It remains to analyse the importance of the non-linear interaction, here $\sim \phi^3$. But this can be done by analogy with sub-section~\ref{sec:4.1}: we go over to the
non-equilibrium representation, via (\ref{gl:hyp}), and then obtain as before a $1/t$-contribution in the generalised Schr\"odinger operator $\overline{\mathscr{S}}$, 
in analogy with eq.~(\ref{gl:4.4}). What is different, is that $\overline{\phi}$ has a non-vanishing scaling dimension whenever $b\neq 0$, because of (\ref{18b}). 
Since in sub-section~\ref{sec:4.2} we saw how to treat this, we can conclude that we have the criterion
\BEQ \label{gl:4.25} 
2\xi = 2\left( \frac{\lambda}{\mathpzc{z}}-b\right) > 1
\EEQ
in order that the non-linearity in (\ref{gl:4.24}) should be irrelevant (higher-order non-linearities will be more irrelevant). 

However, the existing bound (\ref{gl:lambda}) 
is not strong enough to permit any definite conclusion.\footnote{Since (\ref{gl:lambda}) at $T=T_c$ is not saturated by the spherical model, 
one might conjecture the existence of an improved estimate. However, the example of the $3D$ Heisenberg model in table~\ref{tab:2}, 
which almost saturates the improved bound (\ref{gl:lambda}) for $T=T_c$, suggests otherwise.} 
Therefore, we content ourselves to give in table~\ref{tab:2} a list of examples of values of $2\xi$, 
for several models in $2D$ and $3D$ quenched onto their critical point $T=T_c$. 
We observe that the criterion (\ref{gl:4.25}) appears to be satisfied in all cases, since $2\xi$ turns out to be larger than unity, 
although by a surprisingly small margin in the case of the $3D$ Heisenberg universality class. 
In the $3D$ diluted Ising model, we allow for a distinction into `strong' and `weak' disorder, as advocated in \cite{Prud15}, 
but do not wish to enter further into this discussion {\it \'epineuse}. 

\begin{table}[tb]
\begin{center}
\begin{tabular}{|l|lllll|l|l|} \hline 
model              & ~$d$~ & ~$\mathpzc{z}$~  & ~$\lambda$~ & ~$\eta$~    & ~$b$~        & ~$2\xi$~    & \\ \hline 
Ising              & 2     & 2.1667           & 1.588       & 0.25        & 0.115        & 1.236       & \\
                   & 3     & 2.055            & 2.78        & 0.0363      & 0.504        & 1.698       & \\[0.2cm]
Potts-3            & 2     & 2.197            & 1.836       & 0.267       & 0.121        & 1.429       & \\[0.2cm]
Potts-4            & 2     & 2.295            & 2.2         & 0.25        & 0.109        & 1.700       & \\[0.2cm]
Turban-3           & 2     & 2.38             & 2.32        & 0.25        & 0.105        & 1.74        & \\[0.2cm]
XY                 & 2     & 2                & 1.48        & 0.25        & 0.125        & 1.23        & \\
                   & 3     & 2                & 2.68        & 0.038       & 0.519        & 1.6         & \\[0.2cm]
Heisenberg         & 3     & 1.975            & 2.05        & 0.038       & 0.526        & 1.02        & \\[0.2cm]
tri-critical Ising & 2     & 2.215            & 3.2         & 0.075       & 0.0339       & 2.82        & \\[0.2cm]
diluted Ising      & 3     & 2.19             & 2.72        & 0.04        & 0.47         & 1.5         & weak \\
                   & 3     & 2.66             & 2.61        & 0.04        & 0.39         & 1.2         & strong \\ \hline
\end{tabular}
\caption[tab2]{Numerical estimates of the control parameter $2\xi$ in spin systems quenched onto criticality. 
The data for the exponents $\mathpzc{z}, \lambda,\eta$ are taken from \cite{Henk10} and references therein, and $b$ is calculated from (\ref{gl:val-b}).
This gives  $2\xi=2\bigl(\frac{\lambda}{\mathpzc{z}}-b\bigr)$. 
We also allow for an eventual distinction between `weak' and `strong' disorder in the $3D$ diluted Ising model, see \cite{Prud15}. 
\label{tab:2} }
\end{center}
\end{table}

{\it In spin systems quenched onto $T=T_c$, with local interaction and non-conserved model-A type dynamics and if the criterion (\ref{gl:4.25}) is obeyed, 
the cubic non-linearity in the equation of motion (\ref{gl:4.24}) should be irrelevant for the
long-time dynamical symmetry.} 

A specific comment concerns the phase-ordering kinetics in non-conserved Ising model with power-law interactions of the form $J(r)\sim r^{-d-\sigma}$ 
which for $0<\sigma<1$ leads to a new universality class different from the one with short-ranged interactions, recovered for $\sigma>1$. 
In the $2D$ case, for quenches into $0<T<T_c$ such that $b=0$, one has  $\mathpzc{z}=1+\sigma$ \cite{Chris19,Agra21} and $\lambda=1$ \cite{Chris20} which satisfies (\ref{gl:4.25}). 
In the $1D$ case, where $b=0$, $\mathpzc{z}=1+\sigma$ and $\lambda=\frac{1}{2}$ \cite{Corb19a,Corb19b}, eq.~(\ref{gl:4.25}) is not obeyed. 

Having seen that the effective long-time behaviour should be described by the generalised linear Schr\"odinger operator $\overline{\mathscr{S}}$, we can use
the local scale-invariance ({\sc lsi}) with the generators $X_{\pm 1,0}$, extended to an arbitrary dynamical exponent $\mathpzc{z}$, 
in order to fix the two-time auto-response $R(t,s)=\langle \overline{\phi}(t,\vec{0})\overline{\wit{\phi}}(s,\vec{0})\rangle$. The calculation proceeds formally as before 
and we have, in analogy with (\ref{gl:4.7}), up to an overall normalisation  \cite{Henk06} 
\BEQ \label{gl:4.26}
R(t,s) = \left\langle \overline{\phi}(t,\vec{0}) {\overline{\wit{\phi}}}(s,\vec{0})\right\rangle 
= s^{-1-a} \left( \frac{t}{s}\right)^{1+a'-\lambda/\mathpzc{z}} \left(\frac{t}{s} -1 \right)^{-1-a'} 
\EEQ
where the exponents $a,a',\lambda$ are identified as
\BEQ
\frac{\lambda}{\mathpzc{z}} = 2\delta -\xi \;\; , \;\; 1+a = 2\delta -\xi - \wit{\xi} \;\; , \;\; a'-a = \xi +\wit{\xi}
\EEQ
and as in sub-section~\ref{sec:4.2}, they can be treated as free and independent parameters. 
We shall briefly review comparisons of (\ref{gl:4.26}) with numerical data or exact solutions in a variety of spin models which may or may not have an equation of motion
obviously related to (\ref{gl:4.24}).  

{\it For non-equilibrium critical dynamics after a quench onto $T=T_c$, for local interactions and non-conserved model-A dynamics but
an arbitrary value of the dynamic exponent $\mathpzc{z}$, validity of the criterion (\ref{gl:4.25}) implies that the two-time auto-response function 
should be given by the form (\ref{gl:4.26}) and where the three exponents $a,a',\lambda$  can be chosen as free and independent parameters.} 

In table~\ref{tab:3} we list examples of systems, quenched onto $T=T_c$ from a totally uncorrelated initial state, 
for which information on the auto-response scaling function is available. 
If no information on $a'-a$ is given, this exponent was not computed in the quoted source(s). This likely will mean that $a'-a$ should be small in these cases. 
{}From table~\ref{tab:3}, the value of $a'-a$ apparently varies considerably between different dynamical universality classes. 

\begin{table}  
\begin{center}
\begin{tabular}{|l|rrrrcl|} \hline 
model & \multicolumn{1}{c}{$d$} & $a$~~       & $a'-a$~~                   & $\lambda/\mathpzc{z}$~~ &  & Ref.  \\\hline
random walk      &              & -1          & 0                          & 0                       &  & \cite{Cugl94} \\[0.2cm] 
OJK-model        &              & $(d-1)/2$   & $-1/2$                     & $d/4$                   &  & \cite{Bert99,Maze04,Henk05a}\\[0.2cm] 
Glauber-Ising    & 1            & 0           & $-1/2$                     & $1/2$                   &  & \cite{Godr00a,Lipp00}\\ 
                 & 2            & $0.115$     & $-0.187(20)$               & $0.732(5)$              &  & \cite{Plei05,Henk06}\\ 
                 & 3            & $0.506$     & $-0.022(5)$                & $1.36$                  &  & \cite{Plei05,Henk06} \\
                 & 3            & $0.519$     & $-0.015$                   & $1.36$                  &  & \cite{Sast25} \\
                 & $4-\eps$     & $1-\frac{\vep}{2}$ & O($\eps^2)$         & $2-\frac{7}{12}\vep$    &  O($\eps$) & \cite{Cala02b,Cala02c} \\[0.2cm] 
XY               & 3            & 0.52        &                            & $1.34(5)$               &  & \cite{Abri04} \\                 
                 & $4-\eps$     & $1-\frac{\vep}{2}$ & O($\eps^2)$         & $2-\frac{3}{5}\vep$     &  O($\eps$) & \cite{Cala02b,Cala02c} \\[0.2cm] 
spherical        & $<4$         & $d/2-1$     & 0                          & $3d/4-1$                &  & \cite{Godr00b} \\
	             & $>4$         & $d/2-1$     & 0                          & $d/2$                   &  & \cite{Godr00b} \\[0.2cm]
diluted Ising    & $3$          & $0.40$      &                            & $1.05(3)$               &  & \cite{Sche05,Sche06} \\
                 & $4-\eps$     & $1-\frac{1}{2}\left({\frac{6\eps}{53}}\right)^{\demi}$  
                                              & 0 & $2-\left({\frac{6\eps}{53}}\right)^{\demi}$      & O($\eps^{1/2}$)
                                                                                                        & \cite{Cala02a,Sche06}\\[0.2cm]
{\sc fa}         & $1$          & $1$         & $-3/2$                     & $2$                     &  & \cite{Maye03,Maye04c}\\ 
                 & $>2$         & $1+d/2$     & $-2$                       & $2+d/2$                 &  & \cite{Maye03} \\[0.2cm]
Ising spin glass & 3            & $0.060(4)$  & $-0.76(3)$                 & $0.38(2)$               &  & \cite{Henk05a,Henk05b} \\[0.2cm] 
contact process  & 1            & $-0.681$    & $+0.270(10)$               & $1.76(5)$               &  & \cite{Enss04a,Hinrichsen06a,Henk06} \\
                 & $4-\eps$     & $1-\frac{\eps}{2}$ & $0.081(2)\eps$      & $4-\frac{2}{3}\eps$     & O($\eps$) & \cite{Baum07b} \\ 
                 & $>4$         & $d/2-1$     & 0                          & $d/2+2$                 &  & \cite{Ramasco04b} \\[0.2cm] 
{\sc nekim}      & 1            & $-0.430(2)$ & $-0.03(1)$                 & $0.54(2)$               &  & \cite{Odor06a} \\[0.2cm] 
{\sc nepot}-3    & 2            & $0.11$      & $-0.03(3)$                 & $0.815(4)$              &  & \cite{Chat11} \\[0.2cm]
voter Potts-3    & 2            & 0           &                            & $1$                     &  $\log$ & \cite{Chat11} \\[0.2cm]
Kuramoto         &              & -0.6        &                            & 0.34                    &  & \cite{Odor25} \\[0.2cm]
{\sc bcp} = {\sc ew} & $\geq 1$ & $d/2-1$     & 0                          & $d/2$                   &  & \cite{Baum05a,Baum05b,Roethlein06a} \\[0.2cm] 
{\sc bpcp}       & $>2$         & $d/2-1$     & 0                          & $d/2$                   & $\alpha\leq\alpha_C$ & \cite{Baum05a,Baum05b} \\[0.2cm]
{\sc kpz}        & 1            & $-\frac{1}{3}$ & $-\frac{1}{6}$          & $0.667$                 &  & \cite{Henk12} \\
                 & 2            & $0.24(1)$   & $-0.24$                    & $1.24$                  &  & \cite{Kell16,Kell17}\\[0.2cm]
Arcetri          & $<2$         & $d/2-1$     & 0                          & $3d/2-1$                &  & \cite{Henk15}\\
                 & $>2$         & $d/2-1$     & 0                          & $d$                     &  {\small = {\sc ew}} & \cite{Henk15}\\ \hline  
\end{tabular}\end{center}
\caption[tab3]{Systems quenched to a critical point of their stationary state, with the auto-response being described by (\ref{gl:4.26}). 
The numbers in brackets are the uncertainty in the last digit(s). 
Notations are: {\sc fa} Frederikson-Andersen model, {\sc nekim} non-equilibrium kinetic Ising model, {\sc nepot}-3 three-states nonequilibrium Potts model,
{\sc bcp} and {\sc bpcp} the bosonic contact and pair-contact processes (on a part of the critical line only), respectively. 
Non-equilibrium surface growth models: Edwards-Wilkinson ({\sc ew}), Kardar-Parisi-Zhang ({\sc kpz}) and Arcetri models. 
In the Ising spin glass, a bimodal disorder was used. 
Modified and adapted after \cite{Henk08,Henk10}, see also text. 
\label{tab:3}
}
\end{table}

We shall discuss three examples in more detail. 

\noindent
{\bf Example 1.} In the $1D$ Glauber-Ising model quenched to its critical point at $T=0$ (although (\ref{gl:4.5},\ref{gl:4.25}) only hold marginally), 
the exactly known two-time auto-response function is \cite{Godr00a,Lipp00} 
\BEQ
R(t,s) = s^{-1} \left( \frac{t}{s} - 1 \right)^{-1/2} 
\EEQ
Comparing with (\ref{gl:4.26}) of {\sc lsi}, we read off, along with the well-known $\mathpzc{z}=2$
\BEQ
a=0 \;\; , \;\; a' = -\demi \;\; , \;\; \frac{\lambda}{2} = \demi ~~\Longleftrightarrow~~ \delta = \frac{1}{4} \;\; , \;\; \xi = 0 \;\; , \;\; \wit{\xi} = -\demi
\EEQ
The change of representation (\ref{gl:hyp}) is necessary to describe the exact result in this model. This is an exactly solved example where $a$ and $a'$ are different. 
In addition, the exact space-dependent part of $R(t,s;r)$ has indeed the expected gaussian form (\ref{gl:4.7}) \cite{Godr00a,Lipp00}. 

\noindent
{\bf Example 2.} The contact process universality class is described by Reggeon field theory. Dropping a (non-universal) normalisation constant, 
the one-loop $\vep$-expansion gives for the global auto-response function \cite{Baum07b}
\BEA
\wht{R}(t,s) &=& \bigl(t-s\bigr)^{-1-a+d/\mathpzc{z}} \left(\frac{t}{s}\right)^{1+a-\lambda/\mathpzc{z}} {\cal F}_R\left(\frac{s}{t}\right) 
\;\; , \;\; {\cal F}_R(v) = 1 +\vep \mathfrak{f}(v) 
\nonumber \\
\mathfrak{f}(v) &=&  1 + \frac{v}{12} + \left( \frac{1}{v} -1 \right)\ln\bigl(1-v\bigr) -\demi\text{Li}_2(v)
\EEA
up to terms of order ${\rm O}\bigl(\vep^2\bigr)$. This must be compared to the {\sc lsi}-prediction
\BEA
\wht{R}(t,s) &=& \bigl(t-s\bigr)^{-1-a+d/\mathpzc{z}} \left(\frac{t}{s}\right)^{1+a-\lambda/\mathpzc{z}} {\cal F}_R\left(\frac{s}{t}\right) \;\; , \;\;
{\cal F}_R(v) = \left( 1 - v\right)^{a-a'}  \simeq  1 +\vep \mathfrak{g}(v) 
\nonumber \\
\mathfrak{g}(v) &\simeq& c_1 \ln\bigl(1-v\bigr)  
\EEA
where $a'-a=\vep c_1$ and one restricts to the leading order in $\vep$. 
In the region $t/s\gtrsim 0.1$, a fit produces $c_1=0.081(2)$ \cite{Baum07b}. 
Alternatively, if one sets $c_1=\frac{1}{12}$, the first three terms in the small-$v$ expansions of $\mathfrak{f}(v)$ and $\mathfrak{g}(v)$ agree
such that $\mathfrak{f}(v)-\mathfrak{g}(v)=-\frac{1}{480} v^4 + \ldots \simeq -2\cdot 10^{-3}\, v^4 +\ldots$. This explains the significant improvement of the comparison
with numerical data when $a'-a\ne 0$ is admitted. Remarkably, comparing with the $1D$ numerical data in table~\ref{tab:3}, the one-loop prediction \cite{Baum07b} 
is already quite close to the observed numerical values. 
Very recently, a three-loop calculation of all standard non-equilibrium exponents was carried out in this model \cite{Adzh23}. 

It is well-established, in equilibrium critical phenomena, that the $\vep$-expansion series in the O($n$)-model is merely asymptotic and needs to be re-summed \cite{Zinn21} 
which presently can be done \cite{LeGu85,Guid98} on the basis of perturbative expansions beyond six loops, 
see \cite{Klei99,Komp17,Shal21,Schn23,Henr23,Bona23} and references therein. 
For the O($n$)-model, the dynamical exponent $\mathpzc{z}$ is known from five-loop calculations \cite{Adzh22a,Adzh22b}, 
but other non-equilibrium exponents such as $\lambda$ or $\Theta$, let alone scaling functions, 
have apparently up to now not been studied beyond the two-loop level \cite{Cala02c,Cala03}. This also means that an unresummed  $\vep$-expansion, truncated at a fixed order, 
should be considered as an approximation itself and cannot be used straightforwardly to test more general ideas.

\noindent
{\bf Example 3.} In the $2D/3D$ Glauber-Ising or XY models (more generally O($n$) model), 
the one-loop renormalised field-theory does agree with the {\sc lsi}-prediction (\ref{gl:4.26}), with $a'-a=0$ \cite{Cala02b}. 
However, a comparison of numerical data with the two-loop field-theoretical calculations \cite{Cala02c} (their longish results are not included in table~\ref{tab:3}), 
with the additional assumption $a'=a=0$, does not work \cite{Plei05}. However, if $a'-a\ne 0$ is taken into account, via the change of representation (\ref{gl:hyp}), 
the numerical data for the scaling function can be reproduced very satisfactorily \cite{Henk06}, see also table~\ref{tab:3}. 
For $d>4$, these models are in the mean-field universality class, with the exponents as listed for the spherical model. 
Again, the $\vep$-expansion should in principle be re-summed to become a numerical tool. 

Similarly, in the diluted Ising model there exist one-loop estimates for the exponents and the global auto-response scaling function which, to the leading order 
${\rm O}\bigl(\vep^{1/2}\bigr)$, agree with (\ref{gl:4.26}) \cite{Cala02a,Sche05}.\\[0.2cm] 

The comparison of the spherical model at $T=T_c$ is analogous to the case $T<T_c$ before, see section~\ref{sec:4.1}, and we just refer to table~\ref{tab:3} for the exponents. 
The comparison for the critical Arcetri model \cite{Henk15} (for $d=1$ it has the same equation of motion as the mean spherical spin glass \cite{Cugl95}) 
is similar. For $d>2$, this is in the same dynamical universality class as the Edwards-Wilkinson growth model \cite{Roethlein06a}. 

New results on the form of the response function might become available soon in the exactly solvable voter model with long-range interactions, 
see \cite{Corb24f} and references therein. 

In conclusion, {\it in a large variety of models quenched onto their critical point, the phenomenological prediction (\ref{gl:4.26}) 
appears to be in good agreement with numerical data.} 

\begin{table}  
\begin{center}
\begin{tabular}{|l|rrrrcl|} \hline tore
model & \multicolumn{1}{c}{$d$} & $a$~~       & $a'-a$~~                   & $\lambda/\mathpzc{z}$~~ &  & Ref.  \\\hline
contact process  & 1            & $-0.681$    & $0.002$                    & $1.79$                  &  & \cite{Henk13a} \\[0.3cm]
KPZ              & 1            & $-\frac{1}{3}$ & $-0.49$                 & $0.667$                 &  & \cite{Henk12} \\[0.05cm]
                 & 2            & $0.24(1)$   & $0.25$                     & $1.24$                  &  & \cite{Kell16,Kell-Dr}\\[0.3cm]
Glauber-Ising    & 2            & $0.115$     & $-0.187(20)$               & $0.732(5)$              &  & \cite{Sast25}\\ 
 \hline  
\end{tabular}\end{center}
\caption[tab4]{Systems quenched onto a critical point of their stationary state, with the auto-response given by (\ref{gl:4.32}). 
The numbers in brackets estimate the uncertainty in the last digit(s). 
\label{tab:4}
}
\end{table}

As a last comment we notice that in analogy with logarithmic conformal invariance, 
one may also consider `logarithmic' extensions of Schr\"odinger-invariance which is obtained by admitting Jordan matrix forms for the scaling dimensions 
$\delta$ and the rapidities $\xi,\wit{\xi}$.  Let it be sufficient here to simply state that
the two-time auto-response function is then predicted to take the form \cite{Henk13a,Henk13b}
\BEQ \label{gl:4.32}
R(ys,s) 
= s^{-1-a}\, y^{1+a'-\lambda/\mathpzc{z}} \bigl(y -1 \bigr)^{-1-a'} \left[ h_0 + g_0 \ln\left( 1 - \frac{1}{y}\right) + f_0 \ln^2\left( 1 - \frac{1}{y}\right) \right]
\EEQ
where $h_0,g_0,f_0$ must be fitted to the data (and for simplicity we dropped all terms which would imply a logarithmic correction to the scaling behaviour itself). 
In table~\ref{tab:4} we collect the exponents of a few spin systems whose two-time auto-response has been successfully
compared to (\ref{gl:4.32}). Then the agreement with the numerical data improves considerably beyond the one already achieved when comparing the same data to the
non-logarithmic form (\ref{gl:4.26}) and now holds for the whole range where the data collapse indicative of dynamical scaling 
holds.\footnote{In practise, it turns out that if
one fixes $f_0=0$, the remaining fits always lead to $\bigl|g_0/h_0\bigr|\ll 1$ so that one is back to the non-logarithmic {\sc lsi}. 
It is therefore necessary to include the $\ln^2$-term in (\ref{gl:4.32}).}  
Comparison with the exponents in table~\ref{tab:3} shows that the apparent values of $a'-a$ may be modified considerably. 
It is a possibility that many critical systems might in the longer run turn out to be described in terms of such logarithmic forms.

\section{Conclusions} \label{sec:5}

Revisiting the generic phenomenology of physical ageing far from equilibrium, one usually encounters a set of expectations, met in almost all systems under study., but 
their possible interrelationships are rarely discussed in the litt\'erature. Here, an unifying perspective was provided by showing that they can be all derived from
a single hypothesis: starting from the infinitesimal generator $X$ of a symmetry at equilibrium, we postulate that one should go over to the new representation 
\BEQ \label{5.1} 
X \mapsto \overline{X} = e^{W(t)}\, X \, e^{-W(t)}
\EEQ
where the choice of $W(t)$ should describe the moving-away of the physical system from its equilibrium state. 
In this work, we have been exclusively studying the consequences of the choice  
\BEQ \label{5.2}
W(t) = \xi \ln t
\EEQ
of the intertwining function $W(t)$, see (\ref{gl:hyp}). This implies that the physical fields
\BEQ
\overline{\phi}(t,\vec{r}) = e^{W(t)} \phi(t,\vec{r}) = t^{\xi}\, \phi(t,\vec{r})
\EEQ
will be characterised by a pair of scaling dimensions $(\delta,\xi)$. 
The choice (\ref{5.2}) is meant as a description of the leading long-time asymptotics of the system in question. 
For example, in those cases where the system's scaling is modified by logarithmic corrections, (\ref{5.2}) will have to be modified. In our approach, it is essential
that time-translation-invariance, at equilibrium generated by $X_{-1}=-\partial_t$, 
is broken `softly' by going over to a transformed form $\overline{X}_{-1}$, rather than breaking it `hard' by suppressing it entirely. 
Then in many cases it becomes possible to reduce the arising scaling functions to functions of a single variable, which allowed
the application of standard phenomenological scaling arguments.

Our {\em postulate} (\ref{5.1},\ref{5.2}), or (\ref{gl:hyp}), has been \BLAU{phenomenologically} successful in the sense that the generic properties of ageing systems, 
which have been considered as {\em folklore} since a long time, 
can be derived from two simple assumptions: namely dynamical scaling and (surprisingly) generalised time-translation-invariance. 
This is the content of the Propositions 1-7 and the various corollaries which immediately follow. 
\BLAU{We have not been able to derive the form (\ref{5.2}) of $W(t)$ and we must} leave open the problem of understanding the physical origin of this prescription. 

The phenomenological consequences are not restricted to merely reproduce what has been known for a long time indeed. 
Namely, we have shown that the validity of the celebrate {\sc jss} scaling relation (\ref{JSS}) can be extended to all temperatures $T\leq T_c$ which should open a new
field of numerical studies via the analysis of global correlators and responses. 
In addition, we have derived a new type of finite-size scaling relations in fully finite systems, 
valid for times when the associated lengths $\ell(t)\gtrsim N$ but $\ell(s)\ll N$. 
This might provide alternative means to extract the values of non-equilibrium exponents.

In carrying this out, auxiliary assumptions made (in the sense of \cite{Laka84}) include
\begin{enumerate}
\item simple ageing with an algebraically growing length scale $\ell(t)\sim t^{1/\mathpzc{z}}$. 
\item all initial correlations were assumed short-ranged. Otherwise, results such as $\lambda_C=\lambda_R$ need no longer hold true \cite{Bray91,Bray94a,Pico02,Dutt08}. 
\item attention was restricted to the sole intertwining operator $W(t)=\xi\ln t$, see eq.~(\ref{gl:hyp}). 
\item single-time correlators are not considered here. 
\end{enumerate}
Future work should relax or further clarify at least some of these assumptions. 

Finally, we could also obtain a simple criterion to appreciate the relevance of the non-linear terms in at least one type of equation of motion. It turned out that in many
cases, the non-linearities in the equations of motion appear to be irrelevant for the discussion of the asymptotic symmetries. 
Of course, non-linearities will retain their importance for explicit calculations of the
exponents in question from first principles, rather than fitting them to the available data, as we have done. 
Our criterion allows for the first time to give an argument on why the hypothesis of Local Scale-Invariance ({\sc lsi}) \cite{Henk92,Henk94,Henk02,Henk06} 
should work in specific systems with non-linear equations of motion, as the examples had shown us since a long time. 
This furnishes a basis for future work in trying to find more extensive applications and to broaden the basis of the theory. 

While we have focused on classical systems, it is a topical question what becomes of this ageing phenomenology in the case of non-equilibrium quantum systems, 
see e.g. \cite{Gage15,Mara15,Giam16,Rugg20,Mede20a,Hali21,Wald21,Defe24} and references therein. \BLAU{It is still unknown what might happen in the context of ageing for
a spontaneously broken time-translation-invariance and/or for time-periodic driving, see \cite{Zale23} for a recent review.} 


\newpage

\appsection{A}{Mathematical properties}

The generalisation of time-translation-invariance is based on a few mathematical properties which are collected here for reference 
in the form of a few {\it lemmata}, which can all be checked through direct and straightforward calculations which we shall not spell out.  
This allows for further generalisations, beyond the case needed here in the main text. 

\noindent
{\bf Lemma A.1.} \cite{Mini12,Henk15b} {\it For the constants $\delta,\gamma$ and the non-constant function $g(z)$, the generators
\BEQ \label{A1}
\ell_n = - z^{n+1} \partial_z - \delta (n+1) z^n - \gamma n z^n -  g(z) z^n
\EEQ
obey the conformal Lie algebra $[\ell_n,\ell_m]=(n-m) \ell_{n+m}$ for $n,m\in\mathbb{Z}$.} 

This gives a non-standard form of the conformal generators $\ell_n$. Herein, $\delta$ is called a scaling dimension and $\gamma$ a rapidity. 

\noindent 
{\bf Lemma A.2.} \cite{Stoi22} {\it The generators (\ref{A1}) can be obtained from the usual conformal generators 
$- z^{n+1}\partial_z-\Delta (n+1) z^n$ through a change of the representation}
\BEQ \label{A2} 
\ell_n = e^{G(z)}\left( - z^{n+1}\frac{\partial}{\partial z}-\Delta(n+1) z^n  \right) e^{-G(z)} \;\; , \;\;
G(z) = \gamma \ln z - \int^{z} \!\D z'\: \frac{g(z')}{z'}
\EEQ
{\it and where $\delta = \Delta -\gamma$.} 

This clarifies the mathematical origin of the new generators $\ell_n$ as being equivalent to a familiar representation of 
the well-known conformal algebra. In the main text, we only need the special case $g(z)=0$. 

\noindent 
{\bf Lemma A.3.} \cite{Henk15b} {\it The co-variant two-point function  of the quasi-primary scaling operators $\phi_i$ of the representation 
(\ref{A2})  of the finite-dimensional algebra $\langle \ell_{\pm 1},\ell_0\rangle$ has the form}
\BEQ
\left\langle \phi_1(z_1)\phi_2(z_2)\right\rangle = \delta_{\Delta_1,\Delta_2} \left( z_1 - z_2\right)^{-2\Delta_1} \Gamma_1(z_1) \Gamma_2(z_2) 
\;\; , \;\;
\Gamma_i(z_i) = z^{\gamma_i} \exp\left[  -\int_1^{z_i} \!\D \zeta\: \frac{g(\zeta)}{\zeta} \right]
\EEQ

Logically, this allows to write non-standard forms of co-variant correlators and similarly for three-point functions and so on. 

The (ortho-)conformal symmetry in two space dimensions usually considered works with a pair $z,\bar{z}$ of variables so that one has
a pair $\ell_n,\bar{\ell}_n$ of commuting families of generators, both constructed according to (\ref{A1}). Similarly, one may construct more 
general representations of the Schr\"odinger(-Virasoro) algebra, following \cite{Mini12}. 

\newpage

\appsection{B}{On the single-time correlator} 

We discuss the scaling of the general two-point function  $C(t,s;\vec{r})=\langle\phi(t,\vec{r})\phi(s,\vec{0})\rangle$ 
in the $t-s\to 0$ limit which would reproduce the single-time correlator $C(t;\vec{r})$ but the generic solution (\ref{gl:loes}) is singular in this limit. 
We change the time coordinates into
\BEQ
\mathfrak{t} = \demi\bigl( t+s\bigr) \;\; , \;\; \vep = \demi \bigl( t - s \bigr)
\EEQ 
Then the conditions (\ref{gl:inv}) take the form, with the identification (\ref{gl:3.1}) for a correlator 
\begin{subequations} \label{gl:B2}
\begin{align} 
\left( - \partial_{\mathfrak{t}}   + \frac{2\xi \mathfrak{t}}{\mathfrak{t}^2-\vep^2} \right) C(\mathfrak{t},\vep;r) &= 0  \label{gl:B2a} \\
\left( - \mathfrak{t}\partial_{\mathfrak{t}} - \vep\partial_{\vep}-\frac{1}{\mathpzc{z}}r\partial_r -2(\delta-\xi)\right) C(\mathfrak{t},\vep;r) &= 0 \label{gl:B2b}
\end{align}
\end{subequations}
In the equal-time limit $\vep\to 0$, the second of these, dynamical scaling eq.~(\ref{gl:B2b}), becomes
\BEQ
\left( - \mathfrak{t}\partial_{\mathfrak{t}} -\frac{1}{\mathpzc{z}}r\partial_r -2(\delta-\xi)\right) C(\mathfrak{t},0;{r}) = 0
\EEQ
with the obvious solution, since then $\mathfrak{t}\to t$ and $\lim_{\vep\to 0} C(\mathfrak{t},\vep;r) = C(t;r)$
\BEQ
C(t;r) = t^{-2(\delta-\xi)} F_C\left(\frac{r}{t^{1/\mathpzc{z}}}\right) = t^{-b} F_C\left(\frac{r}{t^{1/\mathpzc{z}}}\right)
\EEQ
with the scaling function $F_C$ and we re-used (\ref{18}). 
This result, from scale-invariance alone, is in agreement with the general expectations \cite{Bray94a,Cugl03,Henk10,Taeu14} of section~\ref{sec:1}, 
as also summarised in figure~\ref{fig2}ab. 

The difficulty is that eq.~(\ref{gl:B2a}) provides too strong a constraint on the scaling function $F_C=F_C(u)$. Writing $u=r t^{-1/\mathpzc{z}}$ gives 
\BEQ \label{gl:B5}
u F_C'(u) + 2\delta \mathpzc{z} F_C(u) =0 ~~\Longrightarrow~~ F_C(u) = F_C(0) u^{-2\delta \mathpzc{z}}
\EEQ
but such a power-law form is {\em not} in agreement with the available evidence. 

There does not seem to be an easy solution to this difficulty. 
Likely, the requirement of co-variance of the single-time correlator, even under an apparently weak condition as generalised time-translation-invariance, is too 
strong a requirement. In the context of Local Scale-Invariance, see \cite{Henk06}, at least if $\mathpzc{z}=2$, 
the Bargman super-selection rules which follow from combined spatial translation- and Galilei-invariance enforce
that the co-variant contribution to the correlators vanish $C_{\rm cov}=0$, which in eq.~(\ref{gl:B5}) would mean $F_C(0)=0$. 
In fact, only response functions will transform co-variantly under time-space transformations and
correlators must be reconstructed from multi-point response functions \cite{Pico04}. 

In the text, we shall restrict to analyse the properties of two-time correlator with $t>s$ 
and shall return to the open problem of predicting the properties of single-time correlators elsewhere.

\newpage

\appsection{C}{Integrated responses} 

We investigate how the scaling of the auto-response function $R(t,s) = s^{-1-a} f_R(t/s)$, 
with the algebraic asymptotic behaviour (\ref{3.8}) of $f_R(y)$,  
influences the scaling behaviour of integrated responses or susceptibilities
$\chi(t,.) \sim \int \!\D u\: R(t,u)$. We shall derive (\ref{3.10}) in the text. 

The most simple case concerns the so-called {\em intermediate susceptibility} \cite{Corb03a}
\BEQ \label{C1} 
\chi_{\rm int}(t,s) = \int_{s/2}^s \!\D u\: R(t,u) = s^{-a} f_{\rm int}(t/s) \;\; , \;\; 
f_{\rm int}(y) \stackrel{y\gg 1}{\simeq} f_{\infty,{\rm int}}\, y^{-\lambda/\mathpzc{z}}
\EEQ
where $f_{\infty,{\rm int}}$ is a finite constant. This is (\ref{3.10a}). 

This is easily seen by  direct integration \cite{Henk10}
\BEA
f_{\rm int}(y) &=& \lim_{s\to\infty} s^a \int_{s/2}^s \!\D u \: u^{-1-a} f_R\left( \frac{ys}{u}\right) 
\:=\: \int_{1/2}^1 \!\D v\: v^{-1-a} f_R\left( \frac{y}{v}\right) \nonumber \\
&\simeq& f_{\infty,R}\, y^{-\lambda/\mathpzc{z}} \underbrace{~\int_{1/2}^1 \!\D v\: v^{\lambda/\mathpzc{z}-1-a}~}_{=\mbox{\rm ~constant}} \nonumber 
\EEA
where we used in the last step the asymptotics (\ref{3.8}) and can then identify $f_{\infty,{\rm int}}$. 

In principle, we could have replaced $s/2$ in (\ref{C1}) by $\sigma s$, where $\sigma$ is a simple and finite constant such as 
$\sigma=\frac{1}{2}$ or $\frac{1}{3}$. 
At first sight, one might consider taking a limit $\sigma\to 0$ to recover the thermoremanent susceptibility $\chi_{\rm TRM}(t,s)$ 
but then one may encounter a divergency at the lower integration limit.

For the {\em thermoremanent susceptibility} one can show that, with the finite constant $f_{\infty,M}$
\BEQ \label{C2} 
\chi_{\rm TRM}(t,s) = \int_{0}^s \!\D u\: R(t,u) = s^{-a} f_{M}(t/s) \;\; , \;\; 
f_{M}(y) \stackrel{y\gg 1}{\simeq} f_{\infty,M}\, y^{-\lambda/\mathpzc{z}}
\EEQ
either for quenches onto $T=T_c>0$ or else, for quenches to $T<T_c$,  at least if $d\geq 2$.  This is (\ref{3.10b}). 

It is better not to use the above calculation but to proceed as follows, in four steps. 
For definiteness, we now restrict to non-conserved (model-A type) dynamics, assume the validity of (\ref{gl:val-a}) 
and a totally disordered initial state. 
\underline{First}, we write the scaling function $f_M(y)$ as follows
\BEQ \label{C3}
f_M(y) = y^{-a} \int_0^{1/y} \!\D v\: v^{-1-a}\, f_R(1/v)
\EEQ
\underline{Second}, we consider the behaviour of the (non-negative) integrand in (\ref{C3}) for $\frac{1}{v}\ll 1$. This is proportional to 
\BD
v^{-1-a} f_R\left(\frac{1}{v}\right) \sim v^{-1-a+\lambda/\mathpzc{z}} = v^{-(\mathpzc{z}+\mathpzc{z}a-\lambda)/\mathpzc{z}} =: v^{-\vph} 
\ED
and we distinguish two cases. Precisely at $T=T_c$, we have from section~\ref{sec:1} that $a\mathpzc{z}=d-2+\eta$ and
\BD
\vph = \frac{\mathpzc{z}+d-2+\eta - \lambda}{\mathpzc{z}} \leq \frac{\mathpzc{z} + d-2+\eta - \bigl( d-1 +\eta/2\bigr)}{\mathpzc{z}} 
= \frac{\mathpzc{z}-1+\eta/2}{\mathpzc{z}} \leq 1 - \frac{1}{2\mathpzc{z}} <1 
\ED
where we used first the critical bound (\ref{gl:lambda}) and then recalled that $\eta\leq 1$ \cite{Simo80} and $\mathpzc{z}<\infty$. 
If otherwise $T<T_c$, we have, again from section~\ref{sec:1}, 
that $\mathpzc{z}a=1$ and
\BD
\vph = \frac{\mathpzc{z}+1-\lambda}{\mathpzc{z}} \leq \frac{\mathpzc{z}+1-d/2}{\mathpzc{z}} 
\stackrel{\scriptsize\cite{Bray94b}}{=} \frac{3-d/2}{2} <1 \mbox{\rm\small ~~;~~if $d>2$}
\ED
where first we used the other bound (\ref{gl:lambda}) and then recall that for non-conserved dynamics $\mathpzc{z}=2$ \cite{Bray94b}.
In both cases, since we have $\vph<1$, the integral in (\ref{C3}) converges at the lower integration limit.  
If $d=2$ in the second case, we recall that the values of $\lambda$ 
are considerably larger than unity\footnote{This holds for short-ranged interactions. 
In the case of long-range power-law interactions, there is for the $2D$ Ising model, 
quenched into $0<T<T_c$ a long-range regime where $\lambda=1$ \cite{Chris20} which must be excluded from the present discussion.}
(actually they are close to $\lambda \approx 1.25$,  see e.g. \cite{Henk10}), 
such that we have once more $\vph<1$ which is needed below for the validity of the mean-value theorem.\footnote{For $d=1$, 
the bound $\lambda\geq d/2$ is too weak to be useful and this case is not discussed here.} 
In spin glasses, where $\mathpzc{z}$ is large, it is enough to notice that $\lambda>1$ for $d>2$ such that $\vph<1$ results once more. 
\underline{Third}, we can conclude that the non-negative function $F(v)$ of the upper integration limit, defined by the integral (\ref{C3}), 
is continuous in the compact interval 
$[0,1/y_{\rm min}]$, with ${1}/{y_{\rm min}}<1$, and differentiable on $(0,1/y_{\rm min}]$ 
such that the mean-value theorem of integral calculus \cite[p. 278]{Hild06} is applicable. 
\underline{Forth}, we can finally estimate the integral using the mean-value theorem 
since $v^{-1-a}f_R(1/v)\geq 0$ can be assumed to be continuous for all $0< \frac{1}{v}\leq \frac{1}{y}<1$. 
With the intermediate value $\bigl(y^*\bigr)^{-1} \in (0,y^{-1})$ or alternatively  
$\bigl( y^*\bigr)^{-1} = c^* y^{-1}$ and $0<c^*<1$ finite, we now also use the asymptotics 
$f_R(y)\simeq f_{\infty,R}\, y^{-\lambda/\mathpzc{z}}$ for $y$ large enough and then have
\BEA
f_M(y) &=& y^{-a} y^{-1} \left( \frac{1}{y^*} \right)^{1+a} f_R\left(\frac{1}{y^*}\right) \nonumber \\
&\simeq& y^{-a-1} \left( \bigl(y^*\bigr)^{-1} \right)^{-1-a} f_{\infty,R}\, \left( \bigl(y^*\bigr)^{-1} \right)^{-\lambda/\mathpzc{z}} 
= f_{\infty,R} \left( c^* \right)^{-1-a+\lambda/\mathpzc{z}} y^{-\lambda/\mathpzc{z}} = \mbox{\rm cste.}\: y^{-\lambda/\mathpzc{z}}
\nonumber 
\EEA
as claimed in (\ref{C2}), and where we identify $f_{\infty,M}$ accordingly. 

Clearly, one may consult \cite{Yeun96a} to extend this argument to the case of conserved dynamics. 

On the other hand, it is not to be advised to try to estimate the zero-field-cooled susceptibility $\chi_{\rm ZFC}(t,s) = \int_s^t \!\D u\: R(t,u)$ 
in this way, since one might neglect dominant contributions to $\chi_{\rm ZFC}$ coming from the upper integration limit 
but not contained in the scaling of $R(t,s)$ \cite{Henk03PP,Henk04PP,Henk10}. 
Systems quenched to $T<T_c$ are particularly sensitive to that 
because the domain walls need not be sharp but their width  may increase as a function of time \cite{Abra89}. 
If that occurs, this effect will dominate the behaviour of $\chi_{\rm ZFC}(t,s)$. 
Since in spin glasses, because of the smallness of the ageing exponent $a$ \cite{Cugl94b}, 
one may often neglect the waiting-time-dependent prefactor $s^{-a}$ in the susceptibility, usually this point needs not be taken into consideration.  

\newpage

\appsection{D}{Finite-size scaling in the spherical model}

We complete the discussion in \cite{Henk23} on finite-size scaling in the kinetic (mean) spherical model by adding information on the response function. 
We also briefly discuss finite-size effects in the $p=2$ fully connected spherical spin glass \cite{Cugl95,Fydo15}. 

The spherical model is formulated in terms of real spins $S_{\vec{n}}(t)\in\mathbb{R}$ attached to sites $\vec{n}\in\Lambda\subset\mathbb{Z}^d$ of a
hyper-cubic lattice. They obey the spherical constraint 
$\sum_{\vec{n}\in\Lambda} \langle S_{\vec{n}}(t)\rangle = |\Lambda| =\prod_{j=1}^d N_j$ which is the number of sites of
$\Lambda$. The hamiltonian is ${\cal H} = -J \sum_{(\vec{n},\vec{m})} S_{\vec{n}} S_{\vec{m}}$ with nearest-neighbour interactions 
and we scale $J\stackrel{!}{=}1$. The model's dynamics follows from the Langevin equation (\ref{2.3a}) \cite{Ronc78,Godr00b} 
\BEQ
\partial_t S_{\vec{n}}(t) = D \Delta_{\vec{n}} S_{\vec{n}}(t) - \mathfrak{z}(t) S_{\vec{n}}(t) + \eta_{\vec{n}}(t) 
\EEQ
with the spatial laplacian $\Delta_{\vec{n}}$. The gaussian white noise $\eta_{\vec{n}}(t)$ has the first two moments
\BD
\langle \eta_{\vec{n}}(t) \rangle = 0 \;\; , \;\; 
\langle \eta_{\vec{n}}(t) \eta_{\vec{m}}(t') \rangle = 2 D T\, \delta(t-t') \delta_{\vec{n},\vec{m}}
\ED
with the temperature $T$ and the kinetic coefficient $D$. With the abbreviations
\BEQ
\omega(\vec{k}) = \sum_{j=1}^d \left( 1 - \cos \frac{2\pi}{N_j} k_j \right) \;\; , \;\;
g(t) = \exp\left( 2 \int_0^{t} \!\D\tau\: \mathfrak{z}(\tau) \right)
\EEQ
the spherical constraint becomes a Volterra integral equation (\ref{2.3b}) for $g(t)$: 
\BEQ \label{D3} 
g(t) = f(t) + 2D\, T \int_0^t \!\D\tau\: g(t) f(t-\tau) \;\; , \;\; f(t) = \frac{1}{|\Lambda|} \sum_{\vec{k}} \exp\bigl( -4D \omega(\vec{k}) t \bigr)
\EEQ
where we abbreviate $\sum_{\vec{k}} := \sum_{k_1=0}^{N_1-1} \cdots \sum_{k_d=0}^{N_d-1}$. We specialise to the hyper-cubic geometry 
$\overbrace{N\times\cdots\times N}^{\mbox{\rm $d^*$ factors}}\times\overbrace{\infty\times\cdots\times\infty}^{\mbox{\rm $d-d^*$ factors}}$ 
and use periodic boundary conditions in the $d^*$ finite directions. Starting from a fully disordered initial state, after a quench to 
$T<T_c(d)$ and for $2<d^*\leq d < 4$ dimensions, solving (\ref{D3}) we have to leading order \cite{Henk23}
\BEQ \label{D4} 
g(t) = \frac{1}{2D T_c} \frac{1}{M_{\rm eq}^2}\, \delta(t) + \frac{\bigl(8\pi D\, t\bigr)^{-d/2}}{M_{\rm eq}^4}\, 
\left( \vartheta_3\left(0,\exp\left[ -\pi \frac{N^2}{8\pi D\, t} \right] \right) \right)^{d^*} 
\EEQ
where $M_{\rm eq}=\bigl( 1 - T/T_c\bigr)^{1/2}$ is the equilibrium magnetisation and 
$\vartheta_3(0,q) =\sum_{n=-\infty}^{\infty} q^{n^2}$ is a Jacobi Theta function \cite{Abra65} which obeys 
$\vartheta_3\bigl(0,e^{-\pi y}\bigr)= y^{-1/2} \vartheta_3\bigl(0, e^{-\pi/y}\bigr)$ and $\vartheta_3\bigl(0,0\bigr)=1$.

A local observable\footnote{It can be shown that fluctuations in $N$ do not modify the leading scaling behaviour of local observables \cite{Anni06}.} 
such as the two-time auto-response function is best evaluated in Fourier space, with the external field $h$ and for $t>s>0$
\BEA
R(t,s) &=& \frac{1}{|\Lambda|} \sum_{\vec{k}} \left. \frac{\delta \langle \wht{S}(t,\vec{k})\rangle}{\delta \wht{h}(s,\vec{k})} \right|_{\wht{h}=0}
\:=\: \frac{1}{|\Lambda|} \sqrt{ \frac{g(s)}{g(t)}\,} \sum_{\vec{k}} e^{2D \omega(\vec{k})(t-s)}
\nonumber \\
&=& \sqrt{\frac{g(s)}{g(t)}\,} \sum_{\vec{q}\in\mathbb{Z}^d} \prod_{j=1}^d e^{-2D(t-s)} I_{N_j q_j}\bigl(2D(t-s)\bigr) \nonumber \\
&\simeq& \sqrt{\frac{g(s)}{g(t)}\,} \frac{1}{(4\pi D(t-s))^{d/2}} 
\prod_{j=1}^{d^*} \sum_{q_j\in\mathbb{Z}} \exp\left[ - \frac{N^2}{4D (t-s)} q_j^2\right] 
\nonumber \\
&=& s^{-d/2} \left( \frac{2 (t/s)^{1/2}}{t/s-1} \right)^{d/2}  
\left[ \frac{\vartheta_3\bigl(0,\exp(-\pi N^2/s)\bigr)}{\vartheta_3\bigl(0,\exp(-\pi N^2/t)\bigr)}
\vartheta_3\left(0,\exp\bigl(-\pi \frac{2 N^2}{t-s}\bigr)\right)^2 \right]^{d^*/2} ~~~~
\label{D5} 
\EEA
where we used in the second line the identity ($I_n(x)$ is a modified Bessel function \cite{Abra65}) 
\BD
\sum_{k=0}^{N-1} \exp\left(  x \cos\frac{2\pi k}{N} \right) 
= N \sum_{q=-\infty}^{\infty} I_{qN}(x)
\ED
In the third line we specialised to the hyper-cubic geometry and used the asymptotics 
$I_n(x) \simeq \bigl( 2\pi x\bigr)^{-1/2} e^{x-n^2/2x}$ for $x\gg 1$ 
and in the forth line finally inserted (\ref{D4}) for positive times. 
In (\ref{D5}) and below we scale $8\pi D \stackrel{!}{=}1$. We observe that the response (\ref{D5}) is temperature-independent, as it should be. 

%
\begin{figure}[t]
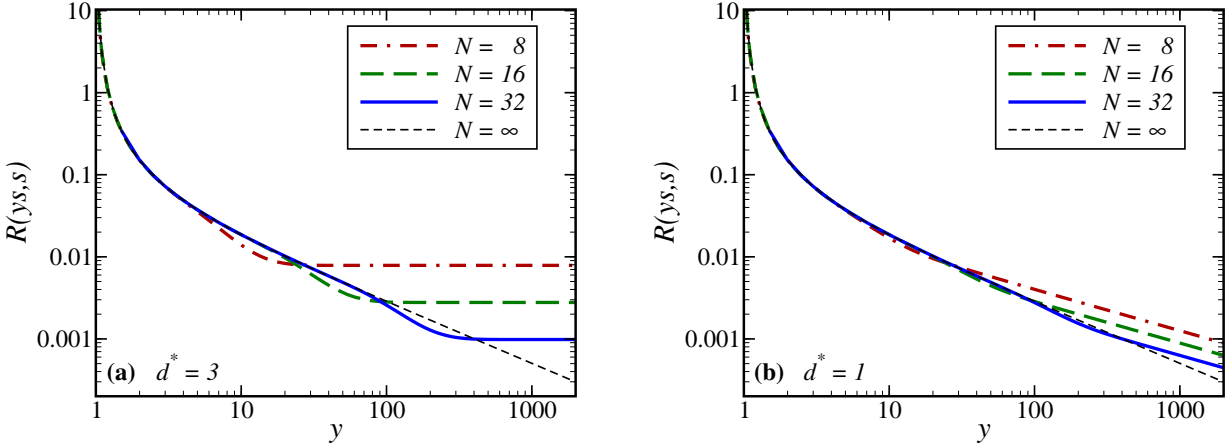

\includegraphics[width=.45\hsize]{TD-symmetrie-RN.eps} ~~~~~ \includegraphics[width=.45\hsize]{TD-symmetrie-RNds1.eps}
\caption[figD1]{Auto-response function of the $3D$ spherical model, for (a) the fully finite case $d^*=3$ and (b) the case of a plaque with $d^*=1$. 
The parameters $N=[8,16,32]$ from top to bottom and  $s=10$ are used.  
The dashed line indicates the infinite-system auto-response, with the asymptotics $R(ys,s)\sim y^{-\lambda/\mathpzc{z}}$. See eq.~(\ref{D7}) for the 
decay of $R$ when $d^*<d$. \label{figD1}}
\end{figure}
%
The auto-response (\ref{D5}) of the $3D$ spherical model is shown in figure~\ref{figD1}a for the fully finite case 
$d^*=3$ and figure~\ref{figD1}b for $d^*=1$. 
In analogy with the schematic auto-correlator of figure~\ref{fig3} in the text, for
sufficiently small values of $y$ the behaviour is very close to the one of the infinite system. 
Then finite-size effects lead to the interruption of ageing and the 
scaling behaviour is broken. The decay with $y$ becomes first more rapid as before and is considerably more pronounced for $d^*=3$ than for $d^*=1$. 
In the fully finite case we see for larger values of $y$ a turn-over towards a plateau of height $R_{\infty}^{(2)}$ 
but for $d^*=1$ only the effective decay exponent
of $R(ys,s)$ is modified, where the amplitude depends on the waiting time $s$. 

%
\begin{figure}[t]
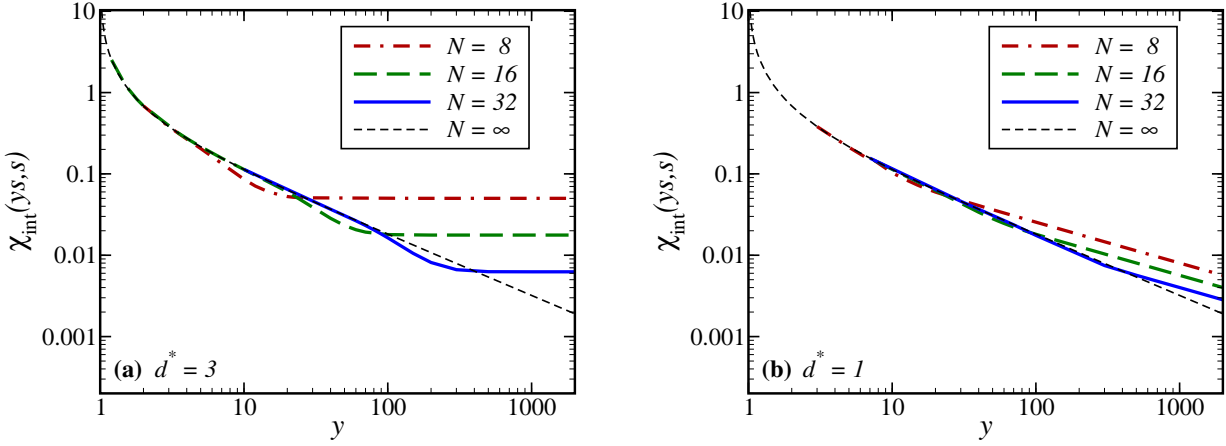

\includegraphics[width=.45\hsize]{TD-symmetrie-chiiN.eps} ~~~~~ \includegraphics[width=.45\hsize]{TD-symmetrie-chiiNds1.eps}
\caption[figD2]{Intermediate integrated response of the $3D$ spherical model, for (a) the fully finite case $d^*=3$ and 
(b) the case of a plaque with $d^*=1$. The parameters $N=[8,16,32]$ from top to bottom and  $s=10$ are used.  
The dashed line indicates the infinite-system integrated response, with the asymptotics $\chi_{\rm int}(ys,s)\sim y^{-\lambda/\mathpzc{z}}$. 
\label{figD2}}
\end{figure}
%
In figure~\ref{figD2} we also show the intermediate integrated response $\chi_{\rm int}$ defined in (\ref{3.10a}), 
for $d^*=3$ as well as for $d^*=1$ and obtained by integrating (\ref{D5}).  Their overall appearance is quite 
similar to the ones of the auto-response function in figure~\ref{figD1}; in particular we notice for the fully finite system the cross-over 
to a plateau of constant height. It is not clear, however, whether this is merely a peculiarity of the spherical model. 

{}From (\ref{D5}), we read off the scaling behaviour of the response in the hyper-cubic geometry. For times such that 
$\ell(t)\approx N$ but $\ell(s)\ll N$, both $\frac{N^2}{t}\ll 1$ and $\frac{N^2}{t-s}\ll 1$. We find
\BEQ \label{D7}
R(t,s;N^{-1}) \sim s^{-d/2} \left( \frac{t}{s}\right)^{-d/4} \left( \frac{N}{t^{1/2}} \right)^{-d^*/2}
\EEQ
in agreement with the generally expected finite-size scaling form (\ref{gl:Rfss}). The last factor in (\ref{D7})
gives the asymptotic finite-size scaling function $\mathscr{F}_R(u)\sim  u^{-d^*/2}$ for $u\ll 1$. 
If furthermore $d=d^*$, the response $R$ converges to a $t$-independent plateau (see figure~\ref{figD1}a), whose height $R_{\infty}^{(2)}$ scales
with $s$ and $N$ as predicted in (\ref{gl:plateauR}) and in agreement with the known values $\lambda=d/2$, $\mathpzc{z}=2$ and $a=\frac{d}{2}-1$.  
There is no plateau for $d^*<d$, see figure~\ref{figD1}b.   

For completeness, we recall the analogous expression for the two-time auto-correlator \cite{Henk23}
\BEQ \label{D6}
C(t,s) = M_{\rm eq}^2 \left( \frac{t^{1/2} s^{1/2}}{(t+s)/2}\right)^{d/2} 
\left( \frac{\vartheta_3\bigl(0, \exp(-\pi\frac{2N^2}{t+s})\bigr)}{\sqrt{\vartheta_3\bigl(0, \exp(-\pi\frac{N^2}{t})\bigr)
\vartheta_3\bigl(0, \exp(-\pi\frac{N^2}{s})\bigr)\,}}\right)^{d^*}
\EEQ
Similarly to the response, from (\ref{D6}) we read off, for $\ell(t)\approx N$ but $\ell(s)\ll N$
\BEQ \label{D8}
C(t,s;N^{-1}) \sim \left( \frac{t}{s} \right)^{-d/4} \left( \frac{N}{t^{1/2}} \right)^{-d^*/2}
\EEQ
in agreement with (\ref{25}) and have $\mathscr{F}_C(u)\sim  u^{-d^*/2}$ for $u\ll 1$. For $d=d^*$ 
there is a plateau whose height $C_{\infty}^{(2)}$ scales with $s$ and $N$ as predicted in (\ref{gl:etf-plateau}) \cite{Henk23}. 

For global correlators, fluctuations in the number of spins $|\Lambda|$ must be taken into account. 
The relevant expressions for the spatially infinite system are given in \cite{Anni06,Anni08}.

We now briefly consider the fully connected spherical model spin glass, 
quenched to temperature $T=0$ from a fully disordered initial state \cite{Cugl95,Fydo15}. 
The hamiltonian is ${\cal H} =-\demi \sum_{n,m} J_{n,m} S_n S_m$ 
(with $n,m=1,\ldots,N$) where $J$ is a symmetric random matrix from the gaussian orthogonal ensemble. 
For an infinite system ($N\to\infty$), the Langevin equation of motion is the same as in the 
$1D$ Arcetri model of growing interfaces \cite{Henk15} and the model is known to
be in the same dynamical universality class as the $3D$ spherical ferromagnet \cite{Cugl95}. 
However, when $N$ is finite, this universality is only kept for
times $t,s\ll t_{\rm cross}\sim N^{2/3}$ \cite{Fydo15}. For larger times, new behaviour occurs. 
If $t\gg N^{2/3}$ and $s\ll N^{2/3}$, the noise-averaged auto-correlator is \cite[eq. (45)]{Fydo15}
\BEQ \label{D9} 
\overline{C(t,s;N^{-1})} \sim \left( \frac{s}{N^{2/3}}\right)^{3/4} = \left( \frac{t}{s} \right)^{-3/4} \left( \frac{N}{t^{3/2}} \right)^{-1/2}
\EEQ
which reproduces the expected scaling behaviour (\ref{25}) of a fully finite system. Asymptotically 
$\mathscr{F}_C(u)\sim  u^{-1/2}$ for $u\ll 1$. This should be compared with 
(\ref{D8}) for the $d=d^*=3$ spherical ferromagnet. In particular, the auto-correlator (\ref{D9}) converges to a plateau
\BEQ
\overline{C(t,s;N^{-1})} ~\longrightarrow~ C_{\infty}^{(2)} \sim \frac{s^{3/4}}{N^{1/2}}
\EEQ
as expected in (\ref{gl:etf-plateau}) and we find\footnote{Since the spin glass has the same equation of motion as the $1D$ Arcetri model, we should
expect $\lambda\geq \demi$ \cite{Yeun96a}.} $\lambda=\demi$ and $\mathpzc{z}=\bigl(\frac{3}{4}\bigr)^{-1} \lambda = \frac{2}{3}$. 
These numbers are different from the ones of the spherical ferromagnet which means that the spherical spin glass, for observation times $
t\gtrsim t_{\rm cross}$, is in a new universality class \cite{Fydo15}. The value $\mathpzc{z}=\frac{2}{3}$ 
agrees with the finite-size scaling of $t_{\rm cross}$. 
A similar result holds for the noise-averaged two-time response function \cite[eq. (39)]{Fydo15}
\BEQ \label{D11} 
\overline{R(t,s;N^{-1})} \sim \frac{N^{5/6}}{s^{3/4}\, t^2} 
= s^{-3/2} \left( \frac{t}{s} \right)^{-3/4} \left( \frac{N}{t^{3/2}} \right)^{5/6}
\EEQ
This has once more the form anticipated in (\ref{gl:Rfss}), again with $\mathpzc{z}=\frac{2}{3}$, 
and asymptotically $\mathscr{F}_R(u)\sim  u^{5/6}$ for $u\ll 1$,
but there is no plateau. The scaling (\ref{D11}) should be compared with the response (\ref{D7}) of the spherical ferromagnet. 
Note that the exponents $b$, $a$ and $\lambda/\mathpzc{z}$ in (\ref{D9},\ref{D11}) retain the same values as for the infinite system 
$N=\infty$, as it should be. 

When both times are getting large and $\gtrsim t_{\rm cross}$, 
the domain sizes saturate and the observables become time-translation-invariant \cite{Fydo15}. 

~\\[0.2cm]
\noindent
{\bf Acknowledgements:} It is a pleasure to thank S. Stoimenov, W. Janke, F. Sastre, P. Sollich, C. Chatelain, D. Warkotsch 
for interesting discussions or correspondence. 
This work was supported by the french ANR-PRME UNIOPEN (ANR-22-CE30-0004-01) and by PHC RILA (Dossier 51305UC).


\end{document}